\documentclass[seceq]{ptptex}
\usepackage{wrapft}
\usepackage{graphicx}

\newcommand{\bra}[1]{\langle {#1} |}     
\newcommand{\ket}[1]{| {#1} \rangle}     
\newcommand{\kket}[1]{| {#1} \rangle\!\rangle}     
\newcommand{\rket}[1]{| {#1} )}     
\newcommand{\wtilde}[1]{\widetilde{#1}} 

\newcommand{\ovl}[1]{\overline{#1}}

\def\beq{\begin{eqnarray}}
\def\eeq{\end{eqnarray}}
\def\bsub{\begin{subequations}}
\def\esub{\end{subequations}}
\def\b{\begin{equation}}

\title{
Background of the $su(2)$-Algebraic Many-Fermion Models\\
in the Boson Realization
}
\subtitle{
Construction of Minimum Weight States by Means of\\
an Auxiliary 
$su(2)$-Algebra and its Related Problems
}    

\author{
Yasuhiko {\sc Tsue},$^{1}$ 
Constan\c{c}a {\sc Provid\^encia},$^{2}$ 
Jo\~ao da {\sc Provid\^encia}$^{2}$ and 
Masatoshi {\sc Yamamura}$^{3}$  
}

\inst{
$^{1}$Physics Division, Faculty of Science, Kochi University, Kochi 780-8520, 
Japan\\
$^{2}$Departamento de F\'{i}sica, Universidade de Coimbra, 3004-516 Coimbra, 
Portugal\\
$^{3}$Department of Pure and Applied Physics, 
Faculty of Engineering Science,\\
Kansai University, Suita 564-8680, Japan
\\

}

\recdate{
\today
}

\abst{
The $su(2)$-algebraic many-fermion model is formulated so as to be able to get the unified understanding 
of the structures of three simple models: 
the single-level pairing, the isoscalar proton-neutron pairing and the two-level Lipkin model. 
Basic idea is to introduce an auxiliary $su(2)$-algebra, any generator of which commutes with any 
generator of the starting $su(2)$-algebra. 
With the aid of this algebra, the minimum weight states are completely determined 
in a simple forms. 
Further, concerning the two algebras, boson realization is presented. 
Through this formulation, the behavior of the total fermion 
in the Lipkin model is notably different of those in the 
other two models. 
As supplementary problem, the boson-fermion realization and the Lipkin model 
in the isovector pairing model are investigated. 
}

\begin{document}

\maketitle

\section{Introduction}

It is well known that, with the aid of boson operators, we can describe 
various phenomena of nuclear and hadron physics, successfully. 
Especially, the studies of microscopic structures of the boson operators 
trace back to the year 1960. 
In this year, Marumori, Arvieu \& Veneroni and Baranger\cite{1} 
proposed independently a theory, which is called 
as the quasi-particle random phase approximation. 
With the aid of this theory, we could understand microscopic structure of the 
boson operators describing the collective vibrational motion 
observed in the spherical nuclei. 
In succession, the success of the theory has stimulated the studies 
of higher order corrections and one of the goals is the boson 
expansion theory: 
Belyaev \& Zelevinsky, Marumori, Yamamura (one of the present authors) \& 
Tokunaga, J. da Provid\^encia (one of the present authors) \& Weneser and Marshalek.\cite{2}
We can find various further studies concerning the boson expansion theory in the 
review article by Klein \& Marshalek.\cite{3} 
Especially, this review concentrated on the boson realization of Lie algebra 
governing many-fermion system under consideration. 
The above is a rough sketch of the boson expansion theory at early stage. 
After these studies, too many papers have been published and it is impossible 
to follow them completely. 
Then, hereafter, we will narrow down the discussion to the 
$su(2)$-algebraic many-fermion model and its boson realization.

We know three simple many-fermion models which obey the $su(2)$-algebra: 
(1) the single-level pairing model,\cite{4} 
(2) the isoscalar proton-neutron pairing model\cite{5} and 
(3) the two-level Lipkin model.\cite{6} 
Hereafter, we abbreviate (1), (2) and (3) as (1) the pairing model, 
(2) the isoscalar pairing model and (3) the Lipkin model. 
The $su(2)$-algebra is composed of three generators 
${\wtilde S}_{\pm,0}$, which are of the bilinear 
forms for the fermion operators and ${\wtilde S}_+$ plays a role of 
block for the orthogonal set built on the minimum weight state. 
Further, each model contains the total fermion number operator ${\wtilde N}$. 
It may be interesting to see that these three models are in the triangle relation 
with one another. 
(i) The isoscalar pairing and the Lipkin model consist of two single-particle levels, 
but the pairing model is treated in one single-particle level. 
(ii) The operators ${\wtilde S}_+$ as the building block in the Lipkin and 
the pairing model are the particle-hole pair and the fermion pair coupled to 
angular momentum $J=0$, respectively. 
But, the proton-neutron pair in the isoscalar pairing model does not couple to 
$J=0$. 
(iii) In the pairing and the isoscalar model, ${\wtilde S}_0$ is a linear function of 
${\wtilde N}$. 
But, the Lipkin model does not contain ${\wtilde N}$ explicitly, in other words, 
${\wtilde S}_0$ and ${\wtilde N}$ are independent 
of each other. 
The above is the triangle relation of the three models.

For the above triangle, (i) and (ii) are not so serious as (iii), because 
(i) and (ii) merely determine the framework of the models. 
As was already mentioned, ${\wtilde S}_0$ is a linear function of 
${\wtilde N}$ in the pairing and the isoscalar pairing model. 
Therefore, since ${\wtilde N}$ commutes with the Hamiltonian which 
is widely adopted, the change of the eigenvalue of ${\wtilde S}_0$ 
automatically leads to the change of the total fermion number $N$. 
In the Lipkin model, ${\wtilde S}_0$ is a linear function of the difference 
between the fermion number operators of the two levels and 
${\wtilde N}$ is the sum of both fermion number operators. 
Therefore, the change of the eigenvalue of ${\wtilde S}_0$ corresponds to the 
change of the difference between the fermion numbers of the two levels. 
The above suggests us that, in the Lipkin model, the information on $N$ is contained 
fully in the minimum weight state, but, we do not know in which form $N$ is contained. 
On the other hand, we know that the minimum weight states in the 
pairing and the isoscalar pairing model depend on the fermion numbers partially 
through the seniority numbers. 
From the above reason, we are forced to reconsider the minimum weight states 
in the $su(2)$models and it may be also interesting to investigate how 
${\wtilde S}_{\pm,0}$ in the boson realization are influenced by the 
minimum weight states. 
In this sense, we must recognize that the $su(2)$-algebraic many-fermion models 
contain still open question in spite of long research history.

Main aim of this paper is to give a possible answer of the above-mentioned question. 
In order to arrive at the goal, we must reformulate the $su(2)$-algebraic many-fermion model 
in rather general scheme. 
We start in preparing many-fermion system which is confined in 
$4\Omega_0$ single-particle states 
($\Omega_0$; integer or half-integer). 
It depends on the model under consideration. 
Following a certain idea which will be mentioned concretely 
in \S 2, we construct the $su(2)$-generators ${\wtilde S}_{\pm,0}$. 
Further, we introduce another $su(2)$-algebra, the generators of which are denoted as 
${\wtilde R}_{\pm,0}$. 
The most important condition in our scheme is that both algebras are 
connected to each other through the commutation relation 
[ any of ${\wtilde R}_{\pm,0}$ , any of ${\wtilde S}_{\pm,0}$ ]$=0$. 
If we follow the idea for constructing ${\wtilde S}_{\pm,0}$ and ${\wtilde R}_{\pm,0}$, 
it can be shown that there does not exist any other $su(2)$-algebra which is independent 
of ${\wtilde R}_{\pm,0}$ and satisfies the above commutation relation. 
Conventionally, the minimum weight state $\rket{m_0}$ is determined through the relations 
${\wtilde S}_-\rket{m_0}=0$ and ${\wtilde S}_0\rket{m_0}=-s\rket{m_0}$. 
In addition to the above, we require the conditions 
${\wtilde R}_-\rket{m_0}=0$ and ${\wtilde R}_0\rket{m_0}=-r\rket{m_0}$. 
Then, 
$({\wtilde R}_+)^{r+r_0}\rket{m_0}\ (-r \leq r_0 \leq r)$ 
is also the minimum weight state for ${\wtilde S}_{\pm,0}$. 
Further, in our scheme, we obtain the relation $s+r=\Omega_0$. 
Usually, $s$ and $2r$ are called as the magnitude of the quasi-spin and the seniority number. 
With the help of the condition we required newly, the minimum weight states can be 
completely derived without any device. 
In the Lipkin model, it can be shown that ${\wtilde R}_0$ is a linear function 
of ${\wtilde N}$ and then, $r_0$ is given as a function of $N$ and we can 
determine the minimum weight state as a function of $N$. 
Since [ any of ${\wtilde R}_{\pm,0}$ , any of ${\wtilde S}_{\pm,0}$ ]=0, 
we can apply the idea of the boson realization to each algebra and 
through the relation $s+r=\Omega_0$, both algebras are coupled with each other. 
As supplementary arguments, we take up two subjects. 
One is related to the boson-quasifermion realization for the $su(2)$-model. 
With the aid of this idea, we can describe many-fermion systems which do not 
obey the $su(2)$-algebra exactly. 
The other is concerned with the Lipkin model obeying the sub-algebra of 
the $so(5)$-algebra which describes the isovector pairing model.\cite{7} 
In this treatment, it is shown that $N$ is in the closed relation to 
the reduced isospin which characterizes the minimum weight state 
of the $so(5)$-algebra.

In \S 2, we present the general scheme of our idea concretely. 
Section 3, 4 and 5 are devoted to applying the general scheme in \S 2 
to the pairing, the isoscalar and the Lipkin model, respectively. 
The difference of the Lipkin model from the other two is clarified. 
In \S 6, the boson-quasifermion realization is formulated in 
rather general form. 
In \S 7, after the $so(5)$-algebra is recapitulated, the Lipkin model is treated 
in the form different from that given in \S 5. 
In \S 8, the concluding remarks are mentioned.

\section{General scheme}

Our description of the $su(2)$-algebraic models starts in giving a 
general scheme. 
We treat many-fermion system which is confined in $4\Omega_0$ single-particle 
states. 
Here, $\Omega_0$ denotes integer or half-integer which depends on the model 
under investigation. 
Since $4\Omega_0$ is an even number, all single-particle states are 
divided into equal parts, $P$ and ${\ovl P}$. 
Therefore, as a partner, each single-particle state belonging to $P$ can find 
a single-particle state in ${\ovl P}$. 
It is natural that we must make rules for finding the partners uniquely. 
We express the partner of the state $\alpha$ belonging to $P$ as ${\ovl \alpha}$ and 
fermion operators in $\alpha$ and ${\ovl \alpha}$ as 
$({\tilde c}_{\alpha}, {\tilde c}_{\alpha}^*)$ and $({\tilde c}_{\ovl \alpha}, {\tilde c}_{\ovl \alpha}^*)$, respectively.

For the above system, we define the following operators: 
\beq\label{2-1}
{\wtilde S}_+=\sum_{\alpha}s_{\alpha}{\tilde c}_{\alpha}^*{\tilde c}_{\ovl \alpha}^*\ , \quad
{\wtilde S}_-=\sum_{\alpha}s_{\alpha}{\tilde c}_{\ovl \alpha}{\tilde c}_{\alpha}\ , \quad
{\wtilde S}_0=\frac{1}{2}\sum_{\alpha}({\tilde c}_{\alpha}^*{\tilde c}_{\alpha}+{\tilde c}_{\ovl \alpha}^*{\tilde c}_{\ovl \alpha})
-\Omega_0\ .
\eeq
The symbol $s_{\alpha}$ denotes real number satisfying 
\beq\label{2-2}
s_{\alpha}^2=1\ .
\eeq
The sum $\sum_{\alpha}$ $(\sum_{\ovl \alpha})$ is carried out in all 
single-particle states in $P$ $({\ovl P})$ and, then, we have 
\beq\label{2-3}
\sum_{\alpha}1=2\Omega_0 \ . \qquad \left( \sum_{\ovl \alpha} 1=2\Omega_0 \ \right)\ .
\eeq
It is easily verified that the operators ${\wtilde S}_{\pm,0}$ form the 
$su(2)$-algebra: 
\bsub\label{2-4}
\beq
[\ {\wtilde S}_+\ , \ {\wtilde S}_-\ ]=2{\wtilde S}_0 \ , \qquad
[\ {\wtilde S}_0\ , \ {\wtilde S}_{\pm}\ ]=\pm {\wtilde S}_{\pm} \ .
\label{2-4a}
\eeq
The Casimir operator ${\wtilde {\mib S}}^2$ is defined as 
\beq
{\wtilde {\mib S}}^2={\wtilde S}_+{\wtilde S}_-+{\wtilde S}_0({\wtilde S}_0-1) \ , \qquad
[\ {\wtilde S}_{\pm,0} \ , \ {\wtilde {\mib S}}^2\ ]=0 \ . 
\label{2-4b}
\eeq
\esub
Usually, for the $su(2)$-algebraic model, the Hamiltonian is expressed in terms of ${\wtilde S}_{\pm,0}$. 
Associating to the above $su(2)$-algebra, we introduce another 
$su(2)$-algebra, the generators of which are defined in the form 
\beq
& &{\wtilde R}_+=\sum_{\alpha}{\tilde c}_{\alpha}^*{\tilde c}_{\ovl \alpha}\ , \quad
{\wtilde R}_-=\sum_{\alpha}{\tilde c}_{\ovl \alpha}^*{\tilde c}_{\alpha}\ , \quad
{\wtilde R}_0=\frac{1}{2}\sum_{\alpha}({\tilde c}_{\alpha}^*{\tilde c}_{\alpha}
-{\tilde c}_{\ovl \alpha}^*{\tilde c}_{\ovl \alpha})\ , 
\label{2-5}
\eeq
\vspace{-0.4cm}
\bsub\label{2-6}
\beq
& &[\ {\wtilde R}_+\ , \ {\wtilde R}_-\ ]=2{\wtilde R}_0 \ , \qquad
[\ {\wtilde R}_0\ , \ {\wtilde R}_{\pm}\ ]=\pm {\wtilde R}_{\pm} \ , 
\qquad\qquad\qquad\ \ 
\label{2-6a}\\
& &{\wtilde {\mib R}}^2={\wtilde R}_+{\wtilde R}_-+{\wtilde R}_0({\wtilde R}_0-1) \ . 
\label{2-6b}
\eeq
\esub
The following relation may be indispensable to understand our idea: 
\beq\label{2-7}
[\ {\rm any\ of\ }{\wtilde R}_{\pm,0}\ , \ {\rm any\ of\ }{\wtilde S}_{\pm,0}\ ]=0 \ .
\eeq
The algebra $({\wtilde R}_{\pm,0})$ plays a role auxiliary to the 
algebra $({\wtilde S}_{\pm,0})$ which must plays a central role for 
describing the dynamics induced by the $su(2)$-Hamiltonian. 
The relation (\ref{2-7}) tells us that the above two algebras seems to be 
independent of each other. 
But, as will be later shown, they are not completely independent. 
Hereafter, at some occasions, we will use the terminologies $S$-spin and $R$-spin for 
$({\wtilde S}_{\pm,0})$ and $({\wtilde R}_{\pm,0})$, respectively. 
As far as the authors know, we have never encountered any investigation based on the 
explicit use of the algebra $({\wtilde R}_{\pm,0})$ for the $su(2)$-algebraic 
many-fermion models. 
In this connection, we must mention that there does not exist 
any $su(2)$-algebra, the generators of which are expressed in terms of 
bilinear form such as $\sum_{\alpha}r_{\alpha}{\tilde c}_{\alpha}^*{\tilde c}_{\ovl \alpha}\ (r_\alpha^2=1)$ 
and commute with those of $({\wtilde R}_{\pm,0})$ defined in the relation (\ref{2-5}).

Now, let us search the orthogonal set produced by the above algebras. 
For this aim, we introduce the following states: 
\bsub\label{2-8}
\beq
\rket{m_0}=\left\{
\begin{array}{ll}
\displaystyle \rket{0} \ , & (r=0) \\
\displaystyle \prod_{i=1}^{2r}{\tilde c}_{{\ovl \alpha}_i}^*\rket{0}\ (=\rket{r;({\ovl \alpha})})\ . &
(2r=1,2,\cdots ,2\Omega_0)\\
\end{array}
\right.
\label{2-8a}
\eeq 
Here, $({\ovl \alpha})$ denotes the configuration 
\beq
({\ovl \alpha})={\ovl \alpha}_1,\ {\ovl \alpha}_2,\ \cdots ,\ {\ovl \alpha}_{2r}\ .
\label{2-8b}
\eeq
\esub
The state $\rket{0}$ is the vacuum of $({\tilde c}_{\alpha},\ {\tilde c}_{\alpha}^*)$ and 
$({\tilde c}_{\ovl \alpha},\ {\tilde c}_{\ovl \alpha}^*)$. 
Beforehand, it may be convenient to specify the rule how to arrange 
the single-particle states in $\rket{m_0}$ appropriately. 
For choosing $({\ovl \alpha})$ for a given value of $r$, 
there exist $(2\Omega_0)!/(2r)!(2\Omega_0-2r)!$ possibilities. 
Then, the set $\{\rket{r;({\ovl \alpha})}\}$ forms an orthogonal set: 
\beq\label{2-9}
(r;({\ovl \alpha})\rket{r';({\ovl \alpha}')}=
\left\{
\begin{array}{ll}
\displaystyle \prod_{i=1}^{2r}\delta_{{\ovl \alpha}_i,{\ovl \alpha}_{i'}}\ , & (r=r') \\
\displaystyle 0 \ . & (r\neq r')\\
\end{array}
\right.
\eeq
Further, the following relations are easily verified: 
\beq\label{2-10}
{\wtilde R}_-\rket{r;({\ovl \alpha})}=0 \ , \qquad
{\wtilde S}_-\rket{r;({\ovl \alpha})}=0 \ , \qquad
\eeq
\vspace{-1.0cm}
\bsub\label{2-11}
\beq
& &{\wtilde R}_0\rket{r;({\ovl \alpha})}=-r\rket{r;({\ovl \alpha})} \ ,
\label{2-11a}\\
& &{\wtilde S}_0\rket{r;({\ovl \alpha})}=-s\rket{r;({\ovl \alpha})} \ . 
\quad (s=\Omega_0-r)
\label{2-11b}
\eeq
\esub
It is important to see that $\rket{r;({\ovl \alpha})}$ 
is a minimum weight state of $R$- and the $S$-spin and the eigenvalues 
of ${\wtilde R}_0$ and ${\wtilde S}_0$ are not independent of each other, 
but restricted to 
\beq\label{2-12}
r+s=\Omega_0 \ .
\eeq
It should be noted that $\Omega_0$ is the most basic parameter in the
present many-fermion system and its value is automatically fixed for a given model. 
Therefore, we cannot choose the values of $r$ and $s$ independently of each other. 
This is the reason why the two algebras are not completely independent 
of each other. 
The eigenstate of ${\wtilde {\mib R}}^2$, ${\wtilde {\mib S}}^2$, ${\wtilde R}_0$ 
and ${\wtilde S}_0$ with the eigenvalues $r(r+1)$, 
$s(s+1)$ $(s=\Omega_0-r)$, $r_0$ and $s_0$ is expressed as 
\bsub\label{2-13}
\beq
& &\rket{s(=\Omega_0-r)\ s_0,rr_0;({\ovl \alpha})}=\left({\wtilde S}_+\right)^{s+s_0}
\rket{rr_0;({\ovl \alpha})} \ , 
\label{2-13a}\\
& &\rket{rr_0;({\ovl \alpha})}=\left({\wtilde R}_+\right)^{r+r_0}\rket{r;({\ovl \alpha})} \ , 
\label{2-13b}\\
& &\qquad r=0,1/2, 1, \cdots , \Omega_0-1/2, \Omega_0 \ , \qquad
r_0=-r, -r+1, \cdots , r-1, r , \nonumber\\
& &\qquad s=\Omega_0-r , \qquad 
s_0=-s, -s+1, \cdots , s-1, s \ .
\label{2-13c}
\eeq
The state given in the relation (\ref{2-13a}) can be rewritten as 
\beq
& &\rket{s(=\Omega_0-r)\ s_0,rr_0;({\ovl \alpha})}=\rket{\Omega_0\sigma,rr_0;({\ovl \alpha})}
=\left({\wtilde S}_+\right)^{\sigma}\rket{rr_0;({\ovl \alpha})} \ , \qquad\ \ 
\label{2-13d}\\
& &\qquad
\sigma=0,\ 1,\ \cdots ,\ 2(\Omega_0-r)\ . 
\label{2-13e}
\eeq
\esub
We can see that the state $\rket{\Omega_0\sigma,rr_0;({\ovl \alpha})}$ is expressed in 
terms of three quantum numbers $(\sigma rr_0)$ except the quantum numbers 
$({\ovl \alpha})$ and the parameter $\Omega_0$. 
Here and hereafter, we omit the normalization constant for any state. 
It should be noted that $\rket{rr_0;({\ovl \alpha})}$ is the minimum weight state for 
the $S$-spin which is deeply connected to the dynamics.

As is clear from the above argument, our idea consists of two steps. 
First is to determine the minimum weight states, in which $({\wtilde R}_{\pm,0})$ 
plays a central role. 
Second is to construct the orthogonal set 
connected with the minimum weight states obtained in the first, 
in which $({\wtilde S}_{\pm,0})$ plays a central role. 
The formalism developed in the above is constructed in the frame of one 
kind of the degree of freedom, 
${\tilde c}_\alpha$, ${\tilde c}_{\alpha}^*$, ${\tilde c}_{\ovl \alpha}$ and 
${\tilde c}_{\ovl \alpha}^*$. 
Then, it may be interesting to describe the present system in terms of two kinds 
of degrees of freedom, in which one is for the first and another is 
for the second. 
For the above idea, the use of the boson realizations of the $su(2)$-algebras 
may be a possible candidate.

The relations (\ref{2-4}), (\ref{2-6}) and (\ref{2-7}) suggest us that the counterparts 
of ${\wtilde R}_{\pm,0}$ and ${\wtilde S}_{\pm,0}$ can be expressed in the 
following form: 
\bsub\label{2-14}
\beq
& &{\hat R}_+={\hat d}_P^*{\hat d}_{\ovl P} \ , \qquad
{\hat R}_-={\hat d}_{\ovl P}^*{\hat d}_{P} \ , \qquad
{\hat R}_0=\frac{1}{2}({\hat d}_P^*{\hat d}_P-{\hat d}_{\ovl P}^*{\hat d}_{\ovl P}) \ , 
\label{2-14a}\\
& &{\hat S}_+={\hat a}^*{\hat b} \ , \qquad
{\hat S}_-={\hat b}^*{\hat a} \ , \qquad
{\hat S}_0=\frac{1}{2}({\hat a}^*{\hat a}-{\hat b}^*{\hat b}) \ . 
\label{2-14b}
\eeq
\esub
Here, $({\hat d}_P, {\hat d}_P^*)$ etc. denote boson operators. 
The above is well known by the name of the Schwinger boson representation of the 
$su(2)$-algebra.\cite{8} 
Under the one-to-one correspondence to the original fermion space, we must construct the 
orthogonal set in the boson space. 
First, we set up the following correspondence: 
\beq\label{2-15}
\rket{0} \sim ({\hat b}^*)^{2\Omega_0}\ket{0} \ .
\eeq
Here, $\ket{0}$ denotes the boson vacuum. 
Next, we notice that the state $\rket{r;({\ovl \alpha})}$ is obtained by 
operating $2r$ fermion creation operators in ${\ovl P}$, i.e., $\prod_{i=1}^{2r}{\tilde c}_{{\ovl \alpha}_i}^*$,  
on the vacuum $\rket{0}$. 
This procedure may be transcribed in the boson space in the following manner: 
The counterpart of $\rket{r;({\ovl \alpha})}$ may be obtained by performing 
$2r$-time operation of $({\hat d}_{\ovl P}^*{\hat b})$ on $({\hat b}^*)^{2\Omega_0}\ket{0}$. 
This is formulated as 
\beq\label{2-16}
\ket{r;{\ovl P}}=
({\hat d}_{\ovl P}^*{\hat b})^{2r}\cdot ({\hat b}^*)^{2\Omega_0}\ket{0}
=({\hat d}_{\ovl P}^*)^{2r}({\hat b}^*)^{2(\Omega_0-r)}\ket{0} \ .
\eeq
Strictly speaking, $\ket{r;{\ovl P}}$ is not counterpart of $\rket{r;({\ovl \alpha})}$, 
because $\ket{r;{\ovl P}}$ does not contain $({\ovl \alpha})$. 
But, the dynamics induced by the $S$-spin depends on the minimum weight state only 
through $r$. 
Later, it will be shown. 
Therefore, at the present framework, it may be not always necessary to make 
$\ket{r;{\ovl P}}$ depend on $({\ovl \alpha})$. 
In \S 6, we will contact again with this point. 
With the aid of the relation (\ref{2-14}), we can prove that 
$\ket{r;{\ovl P}}$ satisfies the same relations as the relation 
(\ref{2-10}) and (\ref{2-11}). 
Further, we have 
\beq\label{2-17}
\ket{rr_0;{\ovl P}}=\left({\hat R}_+\right)^{r+r_0}\ket{r;{\ovl P}}
=({\hat d}_P^*)^{r+r_0}({\hat d}_{\ovl P}^*)^{r-r_0}({\hat b}^*)^{2(\Omega_0-r)}\ket{0} \ . 
\eeq
It may be self-evident that the counterpart of $\rket{s(=\Omega_0-r)\ s_0,rr_0;({\ovl \alpha})}$ 
is obtained in the following from: 
\beq\label{2-18}
& &\ket{s(=\Omega_0-r)\ s_0,rr_0;{\ovl P}}=\left({\hat S}_+\right)^{s+s_0}\ket{rr_0;{\ovl P}} \nonumber\\
&=&({\hat a}^*)^{s+s_0}({\hat b}^*)^{s-s_0}({\hat d}_P^*)^{r+r_0}({\hat d}_{\ovl P}^*)^{r-r_0}\ket{0} \ . 
\quad (s=\Omega_0-r)
\eeq
The state (\ref{2-18}) satisfies the relation (\ref{2-13c}). 
Formally, the eigenstate of the $R$- and the $S$-spin with the eigenvalues $(r,\ r_0)$ and 
$(s,\ s_0)$ can be expressed in the form 
\beq\label{2-19}
& &\ket{ss_0,rr_0}=({\hat a}^*)^{s+s_0}({\hat b}^*)^{s-s_0}({\hat d}_P^*)^{r+r_0}({\hat d}_{\ovl P}^*)^{r-r_0}\ket{0} \ , 
\eeq
If we add the condition (\ref{2-12}), the state (\ref{2-19}) is reduced to the state (\ref{2-18}). 
In this sense, the set 
$\{\ket{ss_0,rr_0};r+s=\Omega_0\}$ forms the physical boson space and the condition (\ref{2-12}) 
holds the key to the solution of our problem. 
The state (\ref{2-19}) under the condition (\ref{2-12}) is specified by three quantum numbers. 
As for the three, it may be interesting to consider which quantum numbers are possible. 
We pay attention to total fermion number which is a constant of motion in the widely 
known $su(2)$-model. 
As can be suggested in the relation (\ref{2-1}), ${\wtilde S}_0$ is connected to 
${\wtilde N}$ in the form 
\bsub\label{2-20}
\beq
{\wtilde S}_0=\frac{1}{2}{\wtilde N}-\Omega_0 \ , \quad {\rm i.e.,}\quad 
s_0=\frac{1}{2}N-\Omega_0 \ .
\label{2-20a}
\eeq
Here, of course, ${\wtilde N}$ and $N$ denote the total fermion number operator and its eigenvalue, 
respectively. 
Another idea is to connect ${\wtilde R}_0$ to ${\wtilde N}$: 
\beq
{\wtilde R}_0=\frac{1}{2}{\wtilde N}-\Omega_0 \ , \quad {\rm i.e.,}\quad 
r_0=\frac{1}{2}N-\Omega_0 \ . 
\label{2-20b}
\eeq
\esub
If we combine the relation (\ref{2-20}) with the condition (\ref{2-12}), the expression (\ref{2-19}) 
becomes the following: 
\bsub\label{2-21}
\beq
& &\ket{\Omega_0N,rr_0}=({\hat a}^*)^{\frac{1}{2}N-r}({\hat b}^*)^{\frac{1}{2}(4\Omega_0-N)-r}
({\hat d}_P^*)^{r+r_0}({\hat d}_{\ovl P}^*)^{r-r_0}\ket{0} \ , 
\label{2-21a}\\
& &\ket{\Omega_0N,ss_0}=({\hat a}^*)^{s+s_0}({\hat b}^*)^{s-s_0}
({\hat d}_P^*)^{\frac{1}{2}N-r}({\hat d}_{\ovl P}^*)^{\frac{1}{2}(4\Omega_0-N)-r}\ket{0} \ . 
\label{2-21b}
\eeq
\esub
The forms (\ref{2-21a}) and (\ref{2-21b}) are based on the relations (\ref{2-20a}) and (\ref{2-20b}), 
respectively. 
The state (\ref{2-21}) will be discussed in \S\S 3, 4 and 5.

Another idea for escaping from the restriction (\ref{2-12}) is to adopt following representation 
for $({\hat S}_{\pm,0})$: 
\beq\label{2-22}
& &{\hat S}_+={\hat A}^*\cdot\sqrt{2{\hat S}-{\hat A}^*{\hat A}} \ , \qquad
{\hat S}_-=\sqrt{2{\hat S}-{\hat A}^*{\hat A}} \ \cdot{\hat A} \ , 
\nonumber\\
& &{\hat S}_0={\hat A}^*{\hat A}-{\hat S} \ . 
\eeq
Here, $({\hat A},\ {\hat A}^*)$ denotes boson operator and ${\hat S}$ is defined as 
\beq\label{2-23}
{\hat S}=\Omega_0-{\hat R} \ , \qquad
{\hat R}=\frac{1}{2}({\hat d}_P^*{\hat d}_P+{\hat d}_{\ovl P}^*{\hat d}_{\ovl P}) \ .
\eeq
As an operator identity, we have 
\beq\label{2-24}
{\hat {\mib R}}^2={\hat R}({\hat R}+1) \ , \qquad
{\hat {\mib S}}^2={\check S}({\check S}+1) \ , \qquad
{\check S}=\frac{1}{2}({\hat a}^*{\hat a}+{\hat b}^*{\hat b}) \ .
\eeq
The relation (\ref{2-24}) tells us that ${\hat R}$ and ${\check S}$ can be regarded as 
the operators expressing the magnitudes of the $R$- and the $S$-spin, respectively, 
if they are independent of each other. 
In order to connect the $S$-spin to the $R$-spin, we adopt the form 
(\ref{2-23}) as the operator for the magnitude of the $S$-spin, which comes from the 
relation (\ref{2-12}). 
If ${\hat S}$ is replaced with $c$-number $s$, the form (\ref{2-20}) becomes the conventional 
Holstein-Primakoff representation.\cite{9} 
In this sense, we take the form (\ref{2-20}) into the Holstein-Primakoff 
representation. 
We can see that the $S$-spin depends only on $r$. 
The counterpart of $\rket{s(=\Omega_0-r)\ s_0,rr_0;({\ovl \alpha})}$ for the 
Holstein-Primakoff representation is given as 
\beq\label{2-25}
& &\kket{\Omega_0\sigma,rr_0}=({\hat A}^*)^{\sigma}({\hat d}_P^*)^{r+r_0}
({\hat d}_{\ovl P}^*)^{r-r_0}\kket{0}\ , 
\qquad
\left(
{\hat A}\kket{0}={\hat d}_P\kket{0}={\hat d}_{\ovl P}\kket{0}=0 \right) 
\nonumber\\
& &\qquad
\sigma=0,\ 1,\ 2,\ \cdots ,\ 2(\Omega_0-r), \qquad 2(\Omega_0-r)=2s \ .
\eeq
This form comes from the form (\ref{2-13}).

The present boson representation seems to be apparently not so new. 
But, in reality, it contains new features. 
On the occasion of investigating the boson realization of the $su(2)$-algebraic many-fermion models, 
the present framework, which consists of the two steps, teaches us that it 
may be necessary to take into account not only the algebra $({\hat S}_{\pm,0})$ but also 
$({\hat R}_{\pm,0})$. 
Especially, it is interesting to investigate how the operator ${\hat R}_0$ influences the results. 
In \S 3 $\sim$ 5, 
we will present the results given in three concrete models in relation to the 
effects of ${\hat R}_0$.

\section{The pairing model}

First example of the application of the general scheme given in \S 2 
is the single-level pairing model.\cite{4} 
This model may be one of most basic models in nuclear physics. 
It consists of identical fermions in one single-particle level 
with the degeneracy $2j+1$ ($=2\Omega$, $j$; half-integer). 
We denote the fermion operators as $({\tilde c}_m,\ {\tilde c}_m^*)$, where 
$m=\pm 1/2,\ \pm 3/2, \cdots ,\ \pm(j-1),\ \pm j$. 
For the above system, we define the following operators: 
\beq\label{3-1}
& &{\wtilde S}_+=\frac{1}{2}\sum_m (-)^{j-m}{\tilde c}_m^*{\tilde c}_{-m}^*\ , \qquad
{\wtilde S}_-=\frac{1}{2}\sum_m (-)^{j-m}{\tilde c}_{-m}{\tilde c}_{m}\ , \nonumber\\
& &{\wtilde S}_0=\frac{1}{2}\sum_m {\tilde c}_m^*{\tilde c}_m -\frac{1}{2}\Omega\ .
\eeq
The operator ${\wtilde S}_+$ plays a role of creation of the Cooper pair and the set 
$({\wtilde S}_{\pm,0})$ forms the $su(2)$-algebra (\ref{2-4}). 
The expression (\ref{3-1}) can be rewritten to 
\beq\label{3-2}
& &{\wtilde S}_+=\sum_{m>0} (-)^{j-m}{\tilde c}_m^*{\tilde c}_{-m}^*\ , \qquad
{\wtilde S}_-=\sum_{m>0} (-)^{j-m}{\tilde c}_{-m}{\tilde c}_{m}\ , \nonumber\\
& &{\wtilde S}_0=\frac{1}{2}\sum_{m>0}( {\tilde c}_m^*{\tilde c}_m+{\tilde c}_{-m}^*{\tilde c}_{-m}) -\frac{1}{2}\Omega\ .
\eeq
The expression (\ref{3-2}) is reduced to the form (\ref{2-1}), if $m$, $-m$, $(-)^{j-m}$ and 
$\Omega$ read, for $m>0$, 
\beq\label{3-3}
m \longrightarrow \alpha\ , \qquad
-m \longrightarrow {\ovl \alpha}\ , \qquad
(-)^{j-m}\longrightarrow s_{\alpha}\ , \qquad
\Omega \longrightarrow 2\Omega_0 \ .
\eeq
Hereafter, at some occasions, we use ${\ovl m}$ for $-m$. 
The form (\ref{3-2}) suggests us that $P$ and ${\ovl P}$ consist of positive $m$ and negative 
$m$, respectively, and the states $m$ and $-m$ are in the relation of the 
partner. 
Practically, this choice of the partner may be unique.

Under the reading (\ref{3-3}), we can define ${\wtilde R}_{\pm,0}$ in the form 
\beq\label{3-4}
{\wtilde R}_+=\sum_{m>0}{\tilde c}_m^*{\tilde c}_{-m} \ , \quad
{\wtilde R}_-=\sum_{m>0}{\tilde c}_{-m}^*{\tilde c}_m \ , \quad
{\wtilde R}_0=\frac{1}{2}\sum_{m>0}({\tilde c}_m^*{\tilde c}_m-{\tilde c}_{-m}^*{\tilde c}_{-m})\ .
\eeq
The operators ${\wtilde R}_{\pm,0}$ obey the $su(2)$-algebra and satisfy the relation (\ref{2-7}). 
In this connection, we know another $su(2)$-algebra, which satisfies the relation (\ref{2-7}): 
\beq\label{3-5}
& &{\tilde j}_+=\sum_m \mu_m(j){\tilde c}_m^*{\tilde c}_{m-1}\ , \quad
{\tilde j}_-=\sum_m \mu_m(j){\tilde c}_{m-1}^*{\tilde c}_m \ , \quad
{\tilde j}_0=\sum_m m{\tilde c}_m^*{\tilde c}_m\ , 
\nonumber\\
& &\mu_m(j)=\sqrt{(j+m)(j-m+1)}\ . 
\eeq
The set $({\tilde j}_{\pm,0})$ is the angular momentum operator. 
The set $({\wtilde S}_{\pm,0})$ is scalar for $({\tilde j}_{\pm,0})$ and both sets commute 
with each other. 
However, as can be seen in the expression (\ref{3-5}), the set $({\tilde j}_{\pm,0})$ is not 
suitable for our present discussion.

Now, let us discuss the seniority scheme which characterizes the 
present model. 
The state introduced in the relation (\ref{2-13b}) can be expressed as 
$\rket{rr_0;({\ovl m})}$ in the present notation. 
Here, $({\ovl m})$ denotes the configuration ${\ovl m}_1$, ${\ovl m}_2$, $\cdots$, 
${\ovl m}_{2r}$. 
It satisfies 
\beq
& &{\wtilde S}_-\rket{rr_0;({\ovl m})}=0 \ , \qquad\qquad \qquad\qquad \qquad\qquad
\label{3-6}
\eeq
\vspace{-0.8cm}
\bsub\label{3-7}
\beq
& &{\wtilde R}_0\rket{rr_0;({\ovl m})}=r_0\rket{rr_0;({\ovl m})} \ , 
\label{3-7a}\\
& &{\wtilde S}_0\rket{rr_0;({\ovl m})}=-s\rket{rr_0;({\ovl m})} \ . \quad
\left( s=\frac{\Omega}{2}-r \right)
\label{3-7b}
\eeq
\esub
The relation (\ref{3-6}) indicates that $\rket{rr_0;({\ovl m})}$ does not contain the 
Cooper pair. 
With the definition of ${\wtilde R}_0$ and ${\wtilde S}_0$, the relation (\ref{3-7}) 
leads us to 
\bsub\label{3-8}
\beq
& &({\wtilde N}_+ +{\wtilde N}_-)\rket{rr_0;({\ovl m})}=2r\rket{rr_0;({\ovl m})} \ , 
\label{3-8a}\\
& &({\wtilde N}_+-{\wtilde N}_-)\rket{rr_0;({\ovl m})}=2r_0\rket{rr_0;({\ovl m})} \ . 
\label{3-8b}
\eeq
\esub
Here, ${\wtilde N}_+$ and ${\wtilde N}_-$ denote the fermion number operators of 
$P$ and ${\ovl P}$, respectively: 
\beq\label{3-9}
{\wtilde N}_+=\sum_{m>0}{\tilde c}_m^*{\tilde c}_m \ , \qquad
{\wtilde N}_-=\sum_{m>0}{\tilde c}_{-m}^*{\tilde c}_{-m} \ . 
\eeq
As was already mentioned, $\rket{rr_0;({\ovl m})}$ does not contain any Cooper pair 
and $({\wtilde N}_+ +{\wtilde N}_-)$ denotes the total fermion number. 
Therefore, we can conclude that $2r$ denotes the number of fermions which do not 
couple to the Cooper pair, that is, the seniority number or the number of the 
unpaired fermion. 
The role of $r_0$ can be interpreted as follows: 
At $r_0=-r$, all unpaired fermions belong to ${\ovl P}$ ($m<0$). 
By successive operation of ${\wtilde R}_+$, the number of the unpaired fermion in $P$ 
increases and passes through the point $r_0=0$ ($2r=$even) or $r_0=\pm 1/2$ 
$(2r=$odd), where the unpaired fermions occupy ${\ovl P}$ and $P$ in equal 
weight. 
Finally, at $r_0=r$, all unpaired fermions belong to $P$ ($m>0$). 
As was mentioned in the above, we can learn how the structure of the minimum weight state 
changes with respect to the increase of $r_0$ from $-r$ to $r$.

By operating ${\wtilde S}_+$ on $\rket{rr_0;({\ovl m})}$ successively, we obtain the state\break 
$\rket{s(=\Omega/2-r)\ s_0,rr_0;({\ovl m})}$. 
The relation (\ref{2-13}) gives us 
\beq\label{3-10}
0 \leq 2r \leq \Omega \ , \qquad
-\left(\frac{\Omega}{2}-r\right) \leq s_0 \leq \frac{\Omega}{2}-r \ .
\eeq
The definition of ${\wtilde S}_0$ shown in the relation (\ref{3-2}) leads us to 
\beq\label{3-11}
s_0=\frac{N}{2}-\frac{\Omega}{2} \ . 
\eeq
Here, $N$ denotes total fermion number in the state $\rket{s(=\Omega/2-r)\ s_0,rr_0:({\ovl m})}$. 
With the use of the relation (\ref{3-10}) and (\ref{3-11}), we have 
\beq\label{3-12}
& &{\rm if}\quad 0\leq N \leq \Omega \ , \qquad 0 \leq 2r \leq N \ , \nonumber\\
& &{\rm if}\quad \Omega\leq N \leq 2\Omega \ , \qquad 0 \leq 2r \leq 2\Omega-N \ . 
\eeq
Under the inequality (\ref{3-12}), automatically, first of the relation (\ref{3-10}) can be 
derived. 
The parameters $\Omega$, $N$ and $2r$ are even or odd numbers and by taking into account this property, 
the relation (\ref{3-12}) gives us the following: 
\bsub\label{3-13}
\beq
& &(1)\ N:{\rm even}, \nonumber\\
& &\qquad 
2r=\left\{
\begin{array}{ll}
0,\ 2,\ \cdots , \ N\ , & (0\leq N <\Omega) \\
0,\ 2,\ \cdots , \ \Omega\ , & (N=\Omega \ , \Omega : {\rm even}) \\
0,\ 2,\ \cdots , \ (2\Omega-N)\ , & (\Omega< N \leq 2\Omega) \\
\end{array}\right.
\label{3-13a}\\
& &(2)\ N:{\rm odd}, \nonumber\\
& &\qquad 
2r=\left\{
\begin{array}{ll}
1,\ 3,\ \cdots , \ N\ , & (1\leq N <\Omega) \\
1,\ 3,\ \cdots , \ \Omega\ , & (N=\Omega \ , \Omega : {\rm odd}) \\
1,\ 3,\ \cdots , \ (2\Omega-N)\ , & (\Omega< N \leq 2\Omega-1) \\
\end{array}\right.
\label{3-13b}
\eeq
\esub
The above is well known rule in the pairing model. 
Thus, we could interpret the seniority scheme in terms of the 
$su(2)$-algebra $({\wtilde R}_{\pm,0})$. 
In this sense, the algebra $({\wtilde R}_{\pm,0})$ may be permitted to call 
the seniority algebra or the seniority spin.

Finally, we must contact with the boson realization for the pairing model. 
Main features have been already discussed in \S 2. 
For the comparison with the other model presented in \S\S 4 and 5, 
only we show the counterpart of $\rket{s(=\Omega/2-r)\ s_0,rr_0;({\ovl m})}$ in the 
Schwinger boson representation in terms of total fermion number $N$. 
The expression (\ref{2-21a}) with $\Omega_0=\Omega/2$ gives us 
\beq\label{3-14}
\ket{\Omega/2\ N,rr_0}
=({\hat a}^*)^{\frac{1}{2}(N-2r)}({\hat b}^*)^{\frac{1}{2}((2\Omega-N)-2r)}
({\hat d}_P^*)^{r+r_0}({\hat d}_{\ovl P}^*)^{r-r_0}\ket{0}\ . 
\eeq
Of course, we used the relation (\ref{3-11}). 
Since $r$ is positive and the exponents of ${\hat a}^*$ and ${\hat b}^*$ should be positive 
or zero, we have $0\leq 2r \leq N$ and $0\leq 2r \leq 2\Omega-N$, which lead us to 
the relation (\ref{3-12}). 
In the present case, the state (\ref{2-25}) is expressed as 
\beq\label{3-15}
& &\kket{\Omega/2\ \sigma,rr_0}
=({\hat A}^*)^\sigma ({\hat d}_P^*)^{r+r_0}({\hat d}_{\ovl P}^*)^{r-r_0}\kket{0} \ , 
\nonumber\\
& &\sigma=0,\ 1,\ 2,\ \cdots ,\ (\Omega-2r)\ . 
\eeq
The total fermion number $N$ is related to $\sigma$ in the following: 
\bsub\label{3-16}
\beq
& &N=2(r+\sigma) \ , \qquad {\rm for}\qquad 
0\leq N \leq \Omega \ , 
\label{3-16a}\\
& &N=2(\Omega-(r+\sigma)) \ , \qquad {\rm for}\qquad 
\Omega \leq N \leq 2\Omega \ . 
\label{3-16b}
\eeq
\esub
The above is the outline of the pairing model based on the general framework in \S 2.

\section{The isoscalar pairing model}

In addition to the pairing model, we know a many-fermion model obeying the $su(2)$-algebra, 
which is called the isoscalar proton-neutron pairing model\cite{5} (in this paper, 
abbreviated as the isoscalar pairing model). 
In this model, two single-particle levels, which we call the $p$- and the 
$n$-level, are occupied by protons and neutrons, respectively. 
The degeneracies are the same as each other: 
$2\Omega=2j+1$ ($j$; half-integer). 
The proton-neutron pairs coupled in the isoscalar type play an central role in this model. 
Of course, the pairs obey the $su(2)$-algebra.

Let us start in giving the isospin operator: 
\beq
& &{\wtilde \tau}_+=\sum_m{\tilde c}_{pm}^*{\tilde c}_{nm} \ , \quad
{\wtilde \tau}_-=\sum_m{\tilde c}_{nm}^*{\tilde c}_{pm} \ , \quad
{\wtilde \tau}_0=\frac{1}{2}\sum_m({\tilde c}_{pm}^*{\tilde c}_{pm} - 
{\tilde c}_{nm}^*{\tilde c}_{nm})  \ , 
\label{4-1}\\
& &[\ {\wtilde \tau}_+\ , \ {\wtilde \tau}_-\ ]=2{\wtilde \tau}_0 \ , \qquad
[\ {\wtilde \tau}_0\ , \ {\wtilde \tau}_{\pm}\ ]=\pm{\wtilde \tau}_{\pm} \ . 
\label{4-2}
\eeq
Here, $m=\pm 1/2,\ \pm 3/2, \ \cdots ,\ \pm (j-1),\ \pm j$. 
For the above isospin operator, we can give the fermion-pair in the isoscalar type ${\wtilde S}_{\pm,0}$: 
\beq\label{4-3}
& &{\wtilde S}_+=\sum_m s_m{\tilde c}_{pm}^*{\tilde c}_{n-m}^* \ , \qquad
{\wtilde S}_-=\sum_m s_m{\tilde c}_{n-m}{\tilde c}_{pm} \ , \nonumber\\
& &{\wtilde S}_0=\frac{1}{2}\sum_m ({\tilde c}_{pm}^*{\tilde c}_{pm}+{\tilde c}_{nm}^*{\tilde c}_{nm})-\Omega \ .
\eeq
The set $({\wtilde S}_{\pm,0})$ forms the $su(2)$-algebra and commutes with $({\wtilde \tau}_{\pm,0})$ 
under the condition 
\beq\label{4-4}
s_{\ovl m}=s_m\ . \qquad (s_m^2=1)
\eeq
The case $s_{\ovl m}=-s_m$, for example, such as $s_m=(-)^{j-m}$, leads us to the isovector type. 
In the case $s_m=(-)^{j-m}$, $s_m=-1$ and $s_m=+1$ appear alternatively as $m$ increases. 
In the present case, $s_m=-1$ and $s_m=+1$ can be freely chosen and, then, without loss of 
generality, we can set $s_m=+1$ for all $m$. 
Hereafter, we will adopt the expression (\ref{4-3}) with $s_m=+1$.

Concerning the idea for defining the $su(2)$-algebra ${\wtilde R}_{\pm,0}$, we 
have two possibilities:
\beq
& &(1)\ {\rm For}\ m>0,\quad p,m\rightarrow \alpha\ , \quad p,-m\rightarrow {\ovl \alpha}, \quad 
{\rm and}\quad n,m\rightarrow \alpha,\quad n,-m\rightarrow {\ovl \alpha}\ , \nonumber\\
& &\qquad\qquad \qquad\quad\   
s_m(=1) \rightarrow s_\alpha\ , \quad \Omega \rightarrow \Omega_0 \ , 
\label{4-5}\\
& &(2)\ {\rm For\ all}\ m,\quad p,m\rightarrow \alpha\ , \quad n,-m\rightarrow {\ovl \alpha}, \quad
s_m(=1) \rightarrow s_\alpha\ , \quad \Omega \rightarrow \Omega_0 \ .
\label{4-6}
\eeq
For the possibility (1), ${\wtilde R}_{\pm,0}$ can be expressed in the form 
\beq\label{4-7}
& &{\wtilde R}_+=\sum_{m>0}
({\tilde c}_{pm}^*{\tilde c}_{p{\ovl m}}+{\tilde c}_{nm}^*{\tilde c}_{n{\ovl m}}) \ , \qquad
{\wtilde R}_-=\sum_{m>0}
({\tilde c}_{p{\ovl m}}^*{\tilde c}_{pm}+{\tilde c}_{n{\ovl m}}^*{\tilde c}_{nm}) \ , 
\nonumber\\
& &{\wtilde R}_0=\frac{1}{2}\sum_{m>0}
\left(({\tilde c}_{pm}^*{\tilde c}_{pm}+{\tilde c}_{nm}^*{\tilde c}_{nm})
-({\tilde c}_{p{\ovl m}}^*{\tilde c}_{p{\ovl m}}+{\tilde c}_{n{\ovl m}}^*{\tilde c}_{n{\ovl m}})\right) \ .
\eeq
However, the form (\ref{4-7}) does not satisfy the relation (\ref{2-7}), which is the most fundamental in our formalism. 
From the above reason, we have to renounce the possibility (1). 
In the possibility (2), ${\wtilde R}_{\pm,0}$ can be expressed as 
\beq\label{4-8}
& &{\wtilde R}_+=\sum_m {\tilde c}_{p{m}}^*{\tilde c}_{n{\ovl m}} \ , \qquad
{\wtilde R}_-=\sum_m {\tilde c}_{n{\ovl m}}^*{\tilde c}_{pm} \ , \nonumber\\
& &{\wtilde R}_0=\frac{1}{2}\sum_m ({\tilde c}_{pm}^*{\tilde c}_{pm}-
{\tilde c}_{n{\ovl m}}^*{\tilde c}_{n{\ovl m}}) \ .
\eeq 
We can prove that the expression (\ref{4-8}) satisfy the relation (\ref{2-7}). 
On the basis of the relation (\ref{4-8}), we will continue our discussion.

In a way similar to the case of the pairing model, we introduce the proton and the 
neutron number operator in the form 
\beq\label{4-9}
{\wtilde N}_p=\sum_m{\tilde c}_{pm}^*{\tilde c}_{pm}\ , \qquad
{\wtilde N}_n=\sum_m{\tilde c}_{n{\ovl m}}^*{\tilde c}_{n{\ovl m}}\ . 
\eeq
Operating ${\wtilde N}_p$ and ${\wtilde N}_n$ on the minimum weight state in the 
present case $\rket{rr_0;(n{\ovl m})}$, we have 
\bsub\label{4-10}
\beq
& &({\wtilde N}_p+{\wtilde N}_n)\rket{rr_0;(n{\ovl m})}=2r\rket{rr_0;(n{\ovl m})}\ , 
\label{4-10a}\\
& &({\wtilde N}_p-{\wtilde N}_n)\rket{rr_0;(n{\ovl m})}=2r_0\rket{rr_0;(n{\ovl m})}\ . 
\label{4-10b}
\eeq
\esub
Here, $(n{\ovl m})$ denotes the configuration $n{\ovl m_1}$, $n{\ovl m_2}$, $\cdots , \ n{\ovl m_{2r}}$. 
We can see that the relation (\ref{4-10}) has the same structure as that 
in the pairing model shown in the relation (\ref{3-8}). 
Therefore, the interpretation given in \S 3 is available without any alternation. 
The quantum number $2r$ indicates the seniority number, the number of the fermions which 
do not couple to the proton-neutron pair in the isoscalar type and it is distributed to 
$(r+r_0)$ protons and $(r-r_0)$ neutrons. 
By operating ${\wtilde S}_+$ successively on the state $\rket{rr_0;(n{\ovl m})}$, we have 
$\rket{s(=\Omega-r)\ s_0,rr_0;(n{\ovl m})}$ and 
operating ${\wtilde N}(={\wtilde N}_p+{\wtilde N}_n)$ on this state, 
we obtain 
the relation 
\beq\label{4-11}
N=2s_0+2\Omega \ . 
\eeq
Noting $-s \leq s_0 \leq s$ and $s=\Omega-r (\geq 0)$, the relation (\ref{4-11}) 
gives us the following inequality: 
\beq\label{4-12}
& &{\rm if}\qquad 0\leq N \leq 2\Omega\ , \qquad 0\leq 2r \leq N\ , \nonumber\\
& &{\rm if}\qquad 2\Omega \leq N \leq 4\Omega\ , \qquad
0\leq 2r \leq 4\Omega-N \ . 
\eeq
The above corresponds to the relation (\ref{3-12}) in the pairing model. 
If discriminating the case $N=$even and the case $N=$odd, the inequality (\ref{4-12}) leads us to 
\bsub\label{4-13}
\beq
& &(1)\ N:{\rm even}, \nonumber\\
& &\qquad 
2r=\left\{
\begin{array}{ll}
0,\ 2,\ \cdots , \ N\ , & (0\leq N \leq2\Omega-2) \\
0,\ 2,\ \cdots , \ 2\Omega\ , & (N=2\Omega) \\
0,\ 2,\ \cdots , \ 4\Omega-N\ , & (2\Omega+2 \leq N \leq 4\Omega) \\
\end{array}\right.
\label{4-13a}\\
& &(2)\ N:{\rm odd}, \nonumber\\
& &\qquad 
2r=\left\{
\begin{array}{ll}
1,\ 3,\ \cdots , \ N\ , & (1\leq N \leq 2\Omega-1) \\
1,\ 3,\ \cdots , \ 4\Omega-N\ , & (2\Omega+1 \leq N \leq 4\Omega-1) \\
\end{array}\right.
\label{4-13b}
\eeq
\esub
It may be interesting to compare the above result with that shown in the relation (\ref{3-13}). 
The quantity $\Omega$ in the pairing model corresponds to $2\Omega$ in the present model 
and, then, even if $\Omega$ is odd, $2\Omega$ is always even. 
This difference appears in the both relations. 
The proton and the neutron number contained in the state $\rket{s=(\Omega-r)\ s_0,rr_0;(n{\ovl m})}$, $N_p$ and $N_n$, are 
given as 
\beq\label{4-14}
N_p=\Omega+s_0+r_0\ , \qquad
N_n=\Omega+s_0-r_0 \ . 
\eeq

In the case $\Omega_0=\Omega$, the counterpart of $\rket{s(=\Omega-r)\ s_0,rr_0;(n{\ovl m})}$ in the 
Schwinger boson representation is obtained from the relation (\ref{2-21}):
\beq\label{4-15}
\ket{\Omega N;rr_0}=({\hat a}^*)^{\frac{1}{2}(N-2r)}
({\hat b}^*)^{\frac{1}{2}((4\Omega-N)-2r)}({\hat d}_P^*)^{r+r_0}({\hat d}_{\ovl P}^*)^{r-r_0}\ket{0} \ . 
\eeq
If $4\Omega$ is replaced with $2\Omega$, the state (\ref{4-15}) becomes the same form as that 
given in the relation (\ref{3-14}). 
The state (\ref{4-15}) gives us the relations $r\geq 0$, $N-2r \geq 0$ and 
$(4\Omega-N)-2r \geq 0$, which are reduced to the relation 
(\ref{4-12}). 
In the present case, the state (\ref{2-25}) becomes 
\beq\label{4-16}
& &\kket{\Omega\sigma,rr_0}=({\hat A}^*)^\sigma ({\hat d}_P^*)^{r+r_0}
({\hat d}_{\ovl P}^*)^{r-r_0}\kket{0}\ , \nonumber\\
& &\sigma=0,\ 1,\ 2,\cdots , \ 2(\Omega-r)\ . 
\eeq
The total fermion number $N$ is related to $\sigma$ in the following: 
\bsub\label{4-17}
\beq
& &N=2(r+\sigma) \ , \qquad {\rm for}\quad 0 \leq N \leq 2\Omega\ , 
\label{4-17a}\\
& &N=2(2\Omega-(r+\sigma)) \ , \qquad {\rm for}\quad 2\Omega \leq N \leq 4\Omega\ . 
\label{4-17b}
\eeq
\esub
The general framework in \S 2 gives us the above result in the case of the isoscalar pairing model.

\section{The Lipkin model}
In this section, we will investigate the model proposed by Lipkin, Meshkov and Glick, which is, 
usually called the Lipkin model.\cite{6}
It aims at schematic description of the particle-hole excitation. 
The Lipkin model consists of two single-particle levels with the same degeneracy 
$2j+1\ (=2\Omega,\ j$:half-integer). 
We discriminate the two levels as the $p$- and the $h$-level. 
The fermion operators in the $p$-and the $h$-level are denoted as 
$({\tilde \gamma}_{pm},\ {\tilde \gamma}_{pm}^*)$ and $({\tilde \gamma}_{hm},\ {\tilde \gamma}_{hm}^*)$, 
respectively, where $m=-j,\ -j+1,\cdots ,\ j-1,\ j$. 
For the above system, we define the following operators: 
\beq
\label{5-1}
& &{\wtilde S}_{+}=\sum_m {\tilde \gamma}_{pm}^*{\tilde \gamma}_{hm}\ , \qquad
{\wtilde S}_{-}=\sum_m {\tilde \gamma}_{hm}^*{\tilde \gamma}_{pm}\ , 
\nonumber\\
& &{\wtilde S}_0=\frac{1}{2}\sum_m ({\tilde \gamma}_{pm}^*{\tilde \gamma}_{pm}-{\tilde \gamma}_{hm}^*{\tilde \gamma}_{hm}) \ .
\eeq
They obey the $su(2)$-algebra (\ref{2-4}). 
Further, we introduce the total fermion number operator ${\wtilde N}$, which is given as 
\beq\label{5-2}
{\wtilde N}=\sum_m ({\tilde \gamma}_{pm}^*{\tilde \gamma}_{pm}+{\tilde \gamma}_{hm}^*{\tilde \gamma}_{hm}) \ .
\eeq
In the pairing and the isoscalar pairing model, ${\wtilde N}$ is expressed as 
${\wtilde N}=2{\wtilde S}_0+2\Omega_0$, in which ${\wtilde N}$ and ${\wtilde S}_0$ depend 
on each other and we have $[\ {\wtilde N}\ , \ {\wtilde S}_{\pm}\ ]=\pm 2{\wtilde S}_\pm (\neq 0)$. 
But, in the Lipkin model, ${\wtilde N}$ is independent of ${\wtilde S}_0$. 
It can be seen in the relations (\ref{5-1}) and (\ref{5-2}) and we have 
$[\ {\wtilde N}\ , \ {\wtilde S}_{\pm}\ ]=0$. 
These points distinguish the Lipkin model from the other two models. 

We introduce the particle and the hole operators $({\tilde c}_{pm},\ {\tilde c}_{pm}^*)$ 
and $({\tilde c}_{hm},\ {\tilde c}_{hm}^*)$ which are related to 
\beq\label{5-3}
{\tilde \gamma}_{pm}={\tilde c}_{pm} \ , \qquad 
{\tilde \gamma}_{hm}=(-)^{j-m}{\tilde c}_{h{\ovl m}}^*\ . 
\eeq
Then, ${\wtilde S}_{\pm,0}$ and ${\wtilde N}$ can be rewritten as 
\beq
& &{\wtilde S}_+=\sum_m(-)^{j-m}{\tilde c}_{pm}^*{\tilde c}_{h{\ovl m}}^* \ , \qquad
{\wtilde S}_-=\sum_m(-)^{j-m}{\tilde c}_{h{\ovl m}}{\tilde c}_{pm} \ , \nonumber\\
& &{\wtilde S}_0=\frac{1}{2}\sum_m({\tilde c}_{pm}^*{\tilde c}_{pm}
+{\tilde c}_{h{\ovl m}}^*{\tilde c}_{h{\ovl m}})-\Omega \ , 
\label{5-4}\\
& &{\wtilde N}=\sum_m({\tilde c}_{pm}^*{\tilde c}_{pm}
-{\tilde c}_{h{\ovl m}}^*{\tilde c}_{h{\ovl m}})+2\Omega \ . 
\label{5-5}
\eeq
The Hamiltonian in the Lipkin model is expressed in terms of the above ${\wtilde S}_{\pm,0}$, 
where ${\wtilde S}_{\pm}$ are expressed in the form of the particle-hole pairs. 
Comparison between the relations (\ref{4-3}) and (\ref{5-4}) is interesting. 
If the index $n$ and the factor $s_m(=+1)$ in the relation (\ref{4-3}) 
read the index $h$ and the factor $(-)^{j-m}$, respectively, the relation (\ref{4-3}) 
becomes identical with the relation (\ref{5-4}). 
Therefore, we can apply the two possibilities (\ref{4-5}) and (\ref{4-6}) for defining 
${\wtilde R}_{\pm,0}$ to the present case: 
\beq
& &(1)\ {\rm For}\ m>0,\quad p,m\rightarrow \alpha\ , \quad p,{\ovl m}\rightarrow {\ovl \alpha}, \quad 
{\rm and}\quad h,m\rightarrow \alpha,\quad h,{\ovl m}\rightarrow {\ovl \alpha}\ , \nonumber\\
& &\qquad\qquad \qquad\quad\   
(-)^{j-m} \rightarrow s_\alpha\ , \quad \Omega \rightarrow \Omega_0 \ , 
\label{5-6}\\
& &(2)\ {\rm For\ all}\ m,\quad p,m\rightarrow \alpha\ , \quad h,{\ovl m}\rightarrow {\ovl \alpha}, \quad
(-)^{j-m} \rightarrow s_\alpha\ , \quad \Omega \rightarrow \Omega_0 \ .
\label{5-7}
\eeq
For the possibility (1), ${\wtilde R}_{\pm,0}$ can be expressed as 
\beq\label{5-8}
& &{\wtilde R}_+=\sum_{m>0}
({\tilde c}_{pm}^*{\tilde c}_{p{\ovl m}}+{\tilde c}_{hm}^*{\tilde c}_{h{\ovl m}}) \ , \qquad
{\wtilde R}_-=\sum_{m>0}
({\tilde c}_{p{\ovl m}}^*{\tilde c}_{pm}+{\tilde c}_{h{\ovl m}}^*{\tilde c}_{hm}) \ , 
\nonumber\\
& &{\wtilde R}_0=\frac{1}{2}\sum_{m>0}
\left(({\tilde c}_{pm}^*{\tilde c}_{pm}+{\tilde c}_{hm}^*{\tilde c}_{hm})
-({\tilde c}_{p{\ovl m}}^*{\tilde c}_{p{\ovl m}}+{\tilde c}_{h{\ovl m}}^*{\tilde c}_{h{\ovl m}})\right) \ .
\eeq
For the possibility (2), ${\wtilde R}_{\pm,0}$ can be expressed as 
\beq\label{5-9}
& &{\wtilde R}_+=\sum_m {\tilde c}_{p{m}}^*{\tilde c}_{h{\ovl m}} \ , \qquad
{\wtilde R}_-=\sum_m {\tilde c}_{h{\ovl m}}^*{\tilde c}_{pm} \ , \nonumber\\
& &{\wtilde R}_0=\frac{1}{2}\sum_m ({\tilde c}_{pm}^*{\tilde c}_{pm}-
{\tilde c}_{h{\ovl m}}^*{\tilde c}_{h{\ovl m}}) \ .
\eeq 
Different from the case of the isoscalar pairing model, we can prove that the relation (\ref{5-8}) and 
(\ref{5-9}) satisfy the relation (\ref{2-7}). 
As will be discussed fully in \S 7, the $su(2)$-algebra for the Lipkin model shown in the relation (\ref{5-4}) 
is a sub-algebra of the $so(5)$-algebra, the typical example of which is the isovector pairing model. 
In the $so(5)$-algebra for the isovector pairing model, only the possibility (1) 
is available. 
Therefore, the possibility (1) in the Lipkin model may be better to discuss in relation to the 
$so(5)$-algebra and from the above-mentioned reason, in this section, we adopt the possibility 
(2) and the configuration $(h{\ovl m})=h{\ovl m}_1,\ h{\ovl m_2},\ \cdots ,\ h{\ovl m_{2r}}$ is used. 
We have already shown that the Lipkin model is analogous to the isoscalar pairing model except 
the treatment of the total fermion number, which is expressed in the form 
\beq\label{5-10}
{\wtilde N}=2{\wtilde R}_0+2\Omega \ . 
\eeq
In the isoscalar pairing model, ${\wtilde N}$ is given as 
\beq\label{5-11}
{\wtilde N}=2{\wtilde S}_0+2\Omega\ . 
\eeq
In \S 2, we have already mentioned that there exist two forms for expressing ${\wtilde N}$ 
in the relations (\ref{2-20a}) and (\ref{2-20b}). 
Certainly, in the Lipkin model, we have the relation (\ref{5-10}) which has been notified in the 
relation (\ref{2-20b}). 
Increases of the total fermion numbers are carried out in terms of the 
successive operations of ${\wtilde R}_+$ and ${\wtilde S}_+$, respectively.

Conventionally, the case of the closed-shell system has been treated in the Lipkin model. 
It corresponds to the case where the $h$-level is fully occupied in the ground state, 
if the interaction is switched off. 
As the interest of physics, it may be acceptable because originally this model was 
proposed with the aim of the schematic understanding of collective 
motion induced by the particle-hole pairs. 
In the framework developed in \S 2, we will discuss the problem how to 
generalize the above-mentioned situation. 
Of course, it may be based on the theoretical interest. 
First, we note the relations 
\bsub\label{5-12}
\beq
& &2{\wtilde S}_0=({\wtilde N}_p+{\wtilde N}_h)-2\Omega\ , 
\label{5-12a}\\
& &2{\wtilde R}_0={\wtilde N}_p-{\wtilde N}_h \ , \qquad {\rm i.e.,}\qquad 
{\wtilde N}=({\wtilde N}_p-{\wtilde N}_h)+2\Omega\ . 
\label{5-12b}
\eeq
\esub
Here, ${\wtilde N}_p$ and ${\wtilde N}_h$ denote 
\beq\label{5-13}
{\wtilde N}_p=\sum_m{\tilde c}_{pm}^*{\tilde c}_{pm}\ , \qquad
{\wtilde N}_h=\sum_m{\tilde c}_{h{\ovl m}}^*{\tilde c}_{h{\ovl m}} \ . 
\eeq
The minimum weight state of the $S$-spin, 
$\rket{rr_0;(h{\ovl m})}$, is the eigenstate 
of ${\wtilde S}_0$ and ${\wtilde R}_0$ with the eigenvalues $-s$ and $r_0$, respectively. 
Therefore, $\rket{rr_0;(h{\ovl m})}$ is the eigenstate of ${\wtilde N}_p$ and ${\wtilde N}_h$, 
the eigenvalues of which are denoted as $n_p$ and $n_h$, respectively. 
The relation (\ref{5-12}) gives as 
\bsub\label{5-14}
\beq
& &2s=2\Omega-(n_p+n_h)\ , 
\label{5-14a}\\
& &N=2\Omega+(n_p-n_h)\ . 
\label{5-14b}
\eeq
\esub
The relation (\ref{5-14}) leads us to 
\beq\label{5-15}
2s=4\Omega-N-2n_h=N-2n_p \ . 
\eeq
Since $\Omega$, $n_p$ and $n_h$ are positive integers and $s$ should be positive, 
the relation (\ref{5-15}) gives us the following: 
\bsub\label{5-16}
\beq
& &(1)\ N:{\rm even}, \nonumber\\
& &\qquad 
2s=\left\{
\begin{array}{ll}
0,\ 2,\ \cdots , \ N\ , & (0\leq N \leq 2\Omega-2) \\
0,\ 2,\ \cdots , \ 2\Omega\ , & (N=2\Omega) \\
0,\ 2,\ \cdots , \ 4\Omega-N\ , & (2\Omega+2 \leq N \leq 4\Omega) \\
\end{array}\right.
\label{5-16a}\\
& &(2)\ N:{\rm odd}, \nonumber\\
& &\qquad 
2s=\left\{
\begin{array}{ll}
1,\ 3,\ \cdots , \ N\ , & (1\leq N \leq 2\Omega-1) \\
1,\ 3,\ \cdots , \ 4\Omega-N\ , & (2\Omega+1 \leq N \leq 4\Omega-1) \\
\end{array}\right.
\label{5-16b}
\eeq
\esub
The above corresponds to the relation (\ref{4-13}). 
On the other hand, with the use of the relations (\ref{5-14b}) and 
$2s \geq 0$ in the relation (\ref{5-15}), 
we derive the result 
\beq
& &0\leq n_h \leq 2\Omega-\frac{N}{2} \ , \qquad
0 \leq n_p \leq \frac{N}{2} \ , 
\label{5-17}\\
& &{\rm if}\quad 0\leq N < 2\Omega\ , \qquad n_p < n_h \ , \nonumber\\
& &{\rm if}\quad N = 2\Omega\ , \qquad n_p = n_h \ , \nonumber\\
& &{\rm if}\quad 2\Omega < N \leq 4\Omega\ , \qquad n_p > n_h \ .
\label{5-18}
\eeq
It may be instructive to draw the relation (\ref{5-17}) and (\ref{5-18}) in figure. 
The closed areas depicted by marks of ``hat (\verb|^|)" (Figs. 1(a) and 1(c)) and the thick line (Fig.1(b)) satisfy the relation 
(\ref{5-17}) and (\ref{5-18}). 
\begin{figure}[t]
\begin{center}
\includegraphics[height=4.8cm]{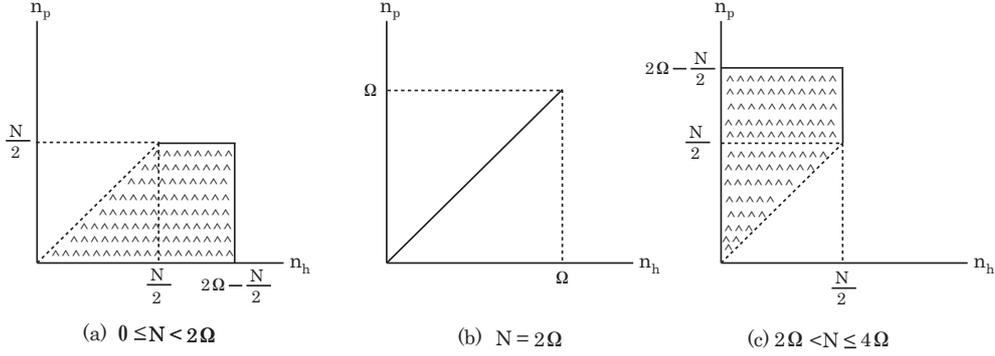}
\caption{The relation (\ref{5-18}) is drawn in (a) 
$0\leq N <2\Omega$, (b) $N=2\Omega$ and (c) $2\Omega < N \leq 4\Omega$, respectively, with 
the relation (\ref{5-17}).
}
\label{fig:1}
\end{center}
\end{figure}

On the basis of the above argument, we will discuss several points concretely. 
As was already mentioned, conventionally, the Lipkin model has been 
treated in the case $N=2\Omega$ and if 
the interaction is switched off, the $h$-level is fully occupied in the ground state. 
This suggests us the relation $n_p=n_h=0$ and the relation (\ref{5-14}) gives us 
$2s=2\Omega$, that is, $2s=2\Omega=N$. 
In relation to this case, the following cases may be interesting: 
Even if $n_h\neq 0$, the case $n_p=0$ leads us to $2s=N$ ($N=2\Omega-n_h <2\Omega)$. 
Further, the case ($n_p\neq 0,\ n_h=0)$ gives us $2s=4\Omega-N$ $(N=2\Omega+n_p>2\Omega)$. 
These cases indicate that $2s$ is expressed only in terms of 
$\Omega$ and $N$. 
However, in other cases, $2s$ is not so simple as that in the above cases. 
For example, if $N=2\Omega$, we have $n_p=n_h(=n_0)$ and 
$2s=N-2n_0$. 
The above argument may be helpful for specifying $2s$ for the Holstein-Primakoff boson realization. 
For this task, the relations (\ref{5-14}) and (\ref{5-15}) are useful.

In the manner similar to the case of the isoscalar pairing model, 
we can express the counterpart of $\rket{s(=\Omega-r)\ s_0,rr_0;(h{\ovl m})}$ in the 
Schwinger boson representation in the following form: 
\beq\label{5-19}
\ket{\Omega N;ss_0}=({\hat a}^*)^{s+s_0}({\hat b}^*)^{s-s_0}
({\hat d}_P^*)^{\frac{1}{2}(N-2s)}({\hat d}_{\ovl P}^*)^{\frac{1}{2}((4\Omega-N)-2s)}\ket{0} \ .
\eeq
The state (\ref{5-19}) gives us the relations $s\geq 0$, $N-2s \geq 0$ and 
$(4\Omega-N)-2s \geq 0$, which are reduced to the relations 
(\ref{5-17}) and (\ref{5-18}). 
Here, we used the relation (\ref{5-14}). 
It may be interesting to see that if $s$ is replaced 
with $r$, the relation (\ref{5-16}) becomes the relation (\ref{4-13}).
The state (\ref{2-25}) in the previous cases is rather different from the present case. 
It may be expressed in the form 
\beq\label{5-20}
& &\kket{\Omega\sigma,rN}=({\hat A}^*)^{\sigma}({\hat d}_P^*)^{r+\frac{1}{2}(N-2\Omega)}
({\hat d}_{\ovl P}^*)^{r-\frac{1}{2}(N-2\Omega)}\kket{0} \ , \nonumber\\
& &\sigma=0,\ 1,\ 2,\ \cdots ,\ 2(\Omega-r) \ .
\eeq 
The total fermion number $N$ is contained in the part of the minimum weight state.

\section{Boson-quasifermion realization}

As a supplementary argument, we consider the boson-fermion realization. 
In \S 2, we presented a possible boson realization of the $su(2)$-algebraic models for 
many-fermion system. 
As can be seen in the relations (\ref{2-14a}) and (\ref{2-25}), at first step, we express the minimum weight states in terms 
of the bosons $({\hat d}_P^*,\ {\hat d}_{\ovl P}^*)$ and, at second step, the orthogonal set constructed on each 
minimum weight state is described in terms of the boson ${\hat A}^*$. 
This point may be interesting, because the two steps are carried out independently or separately from 
each other. 
The use of $({\hat d}_P,\ {\hat d}_P^*,\ {\hat d}_{\ovl P},\ {\hat d}_{\ovl P}^*)$ enables us to specify the minimum weight states 
in terms of $(r,\ r_0)$. 
In the conventional treatment for the $su(2)$-algebraic models, only the $S$-spin is the object of the 
investigation and its Holstein-Primakoff representation is obtained in the 
relation (\ref{2-22}) by replacing ${\hat S}$ with the magnitude of the $S$-spin, $s$. 
But, the task to connect $s$ to the seniority number is performed by each device for each model. 
In our case, without any device, the use of the $R$-spin gives us the relation $s=\Omega_0-r$. 
Further, the $R$-spin orders us to use $r_0$ which may be regarded as one of the 
quantum numbers specifying the minimum weight states. 
The conventional treatment does not contain $r_0$. 
In this sense, compared with the conventional one, 
ours gives us somewhat detailed, but interesting information on the minimum weight states.

In order to get more detailed information, it may be desirable to specify the minimum weight states 
in terms of the quantum numbers $\alpha$, ${\ovl \alpha}$, etc. under the two step scheme. 
For this aim, the following two treatments are instructive: 
1) the boson-quasifermion mapping for the pairing model and 2) the quantization of the 
Dirac bracket appearing in the canonical form of the extended TDHF method 
including the Grassmann variables for the pairing and the Lipkin model. 
The former has been presented by Suzuki and Matsuyanagi\cite{10} and later by 
Hasegawa and Kanesaki\cite{11} and the latter by Kuriyama and one of the present 
authors (M.Y.)\cite{12} 
With the aim of completing the above-mentioned scheme, we introduce the operators 
$({\tilde b}_\alpha,\ {\tilde b}_{\alpha}^*)$ and $({\tilde b}_{\ovl \alpha},\ {\tilde b}_{\ovl \alpha}^*)$ 
governed by the following conditions: 
\bsub\label{6-1}
\beq
& &\{\ {\tilde b}_\alpha\ , \ {\tilde b}_{\beta}^*\ \}
=\delta_{\alpha\beta}-s_\alpha {\tilde b}_{\ovl \alpha}^*(2{\wtilde S})^{-1}s_\beta{\tilde b}_{\ovl \beta} \ , 
\label{6-1a}\\
& &\{\ s_\alpha {\tilde b}_{\ovl \alpha}\ , \ s_\beta {\tilde b}_{\ovl \beta}^*\ \}
=\delta_{\alpha\beta}- {\tilde b}_{\alpha}^*(2{\wtilde S})^{-1}{\tilde b}_{\beta} \ , 
\qquad\qquad
\label{6-1b}
\eeq
\esub
\vspace{-0.6cm}
\bsub\label{6-2}
\beq
& &\{\ {\tilde b}_\alpha\ , \ s_\beta {\tilde b}_{\ovl \beta}^*\ \}
=s_\alpha {\tilde b}_{\ovl \alpha}^*(2{\wtilde S})^{-1}{\tilde b}_{\beta} \ , 
\label{6-2a}\\
& &\{\ s_\alpha {\tilde b}_{\ovl \alpha}\ , \ {\tilde b}_{\beta}^*\ \}
={\tilde b}_{\alpha}^*(2{\wtilde S})^{-1}s_\beta{\tilde b}_{\ovl \beta} \ , 
\qquad\qquad\qquad
\label{6-2b}
\eeq
\esub
\vspace{-0.6cm}
\beq\label{6-3}
\{\ {\tilde b}_\alpha\ , \ {\tilde b}_{\beta}\ \}
=\{\ {\tilde b}_{\ovl \alpha}\ , \ {\tilde b}_{\ovl \beta}\ \}
=\{\ {\tilde b}_\alpha\ , \ {\tilde b}_{\ovl \beta}^*\ \}
=0 \ .
\eeq
Here, $\{\ {\wtilde A}\ , \ {\wtilde B}\ \}$ denotes anti-commutator for 
${\wtilde A}$ and ${\wtilde B}$. 
The operator ${\wtilde S}$ is defined as 
\beq\label{6-4}
{\wtilde S}=\Omega_0-\frac{1}{2}\sum_{\alpha}({\tilde b}_{\alpha}^*{\tilde b}_{\alpha}
+{\tilde b}_{\ovl \alpha}^*{\tilde b}_{\ovl \alpha}) \ .
\eeq
The above anti-commutation relations are closely related to the constraints appearing 
in the canonical form of the constraint system presented by Dirac, that is, the Dirac brackets.\cite{12}

Next, we define ${\tilde \chi}_{\pm}$ in the following bi-linear form: 
\beq\label{6-5}
{\tilde \chi}_+=\sum_{\alpha}s_{\alpha}{\tilde b}_{\alpha}^*{\tilde b}_{\ovl \alpha}^* \ , \qquad
{\tilde \chi}_-=\sum_{\alpha}s_{\alpha}{\tilde b}_{\ovl \alpha}{\tilde b}_{\alpha} \ . 
\eeq
With the use of the relations (\ref{6-1})$\sim$(\ref{6-3}), we can prove the relation 
\beq\label{6-6}
{\tilde b}_{\alpha}{\tilde \chi}_+\ket{0}=s_{\alpha}{\tilde b}_{\ovl \alpha}{\tilde \chi}_+\ket{0}=0 \ . 
\qquad ({\tilde b}_{\alpha}\ket{0}=s_{\alpha}{\tilde b}_{\ovl \alpha}\ket{0}=0)
\eeq
The relation (\ref{6-6}) leads us to 
\beq\label{6-7}
\bra{0}{\tilde \chi}_-\cdot{\tilde \chi}_+\ket{0}=0 \ , \quad
{\rm i.e.,}\quad 
{\tilde \chi}_+\ket{0}=0 \ . 
\eeq
Therefore, in the space spanned by $({\tilde b}_{\alpha}^*, \ {\tilde b}_{\ovl \alpha}^*)$, any state which contains 
${\tilde \chi}_+$ vanishes. 
It indicates that ${\tilde \chi}_+$ play a role of the constraints of the 
above-mentioned Dirac's formalism. 
The relation (\ref{6-3}) tells us that all states constructed by $({\tilde b}_{\alpha},\ {\tilde b}_{\ovl \alpha}^*)$ are 
anti-symmetric with respect to the quantum numbers specifying the single-particle states, 
i.e., fermion type. 
The relation (\ref{6-1}) and (\ref{6-2}) determine the normalization. 
In this sense, ${\tilde b}_{\alpha}$, ${\tilde b}_{\alpha}^*$, ${\tilde b}_{\ovl \alpha}$ and 
${\tilde b}_{\ovl \alpha}^*$ can be called the quasi-fermion operators. 
However, the relation (\ref{6-7}) suggests us that, compared with the original fermion system presented by 
$({\tilde c}_{\alpha}^*,\ {\tilde c}_{\ovl \alpha}^*)$, one degree of freedom is reduced in the system given by 
$({\tilde b}_{\alpha}^*,\ {\tilde b}_{\ovl \alpha}^*)$. 
In order to cancel this discrepancy, we introduce new degree of freedom in terms of boson 
$({\hat A},\ {\hat A}^*)$ satisfying 
\beq\label{6-8}
& &[\ {\hat A}\ , \ {\hat A}^*\ ]=1 \ , \nonumber\\
& &[\ {\tilde b}_{\alpha}\ , \ {\hat A}^*\ ]=[\ {\tilde b}_{\ovl \alpha}\ , \ {\hat A}^*\ ]=
[\ {\tilde b}_{\alpha}\ , \ {\hat A}\ ]=[\ {\tilde b}_{\ovl \alpha}\ , \ {\hat A}\ ]=0 \ . 
\eeq
With the use of ${\tilde b}_{\alpha}$, $s_{\alpha}{\tilde b}_{\ovl \alpha}$, ${\tilde b}_{\alpha}^*$, 
$s_{\alpha}{\tilde b}_{\ovl \alpha}^*$, ${\hat A}$ and ${\hat A}^*$, we define 
the operators ${\tilde c}'_{\alpha}$ and $s_{\alpha}{{\tilde c}'}_{\ovl \alpha}{}^*$ in the form 
\beq\label{6-9}
& &{\tilde c}'_{\alpha}=\sqrt{1-\frac{{\hat A}^*{\hat A}}{2{\wtilde S}}}\ {\tilde b}_{\alpha}
+s_{\alpha}{\tilde b}_{\ovl \alpha}^*\frac{\hat A}{\sqrt{2{\wtilde S}}}\ , \nonumber\\
& &s_{\alpha}{\tilde c}'_{\ovl \alpha}=-{\tilde b}_{\alpha}^*
\frac{\hat A}{\sqrt{2{\wtilde S}}}+
\sqrt{1-\frac{{\hat A}^*{\hat A}}{2{\wtilde S}}}\ s_{\alpha}{\tilde b}_{\ovl \alpha} \ . 
\eeq
We can prove the relation 
\beq\label{6-10}
& &\{\ {\tilde c}'_{\alpha} \ , \ {\tilde c}'_{\beta}{}^*\ \}=
\{\ s_{\alpha}{\tilde c}'_{\ovl \alpha} \ , \ s_{\beta}{\tilde c}'_{\ovl \beta}{}^*\ \}=1 \ , \nonumber\\
& &\{\ {\tilde c}'_{\alpha} \ , \ s_{\beta}{\tilde c}'_{\ovl \beta}{}^*\ \}=
\{\ s_{\alpha}{\tilde c}'_{\ovl \alpha} \ , \ {\tilde c}'_{\beta}{}^*\ \}=0 \ , \nonumber\\
& &\{\ {\tilde c}'_{\alpha} \ , \ {\tilde c}'_{\beta}\ \}=
\{\ {\tilde c}'_{\ovl \alpha} \ , \ {\tilde c}'_{\ovl \beta}\ \}=
\{\ {\tilde c}'_{\alpha} \ , \ {\tilde c}'_{\ovl \beta}\  \}=0 \ . 
\eeq
Here, (and hereafter), we omit the terms related to ${\tilde \chi}_{\pm}$. 
The relation (\ref{6-10}) gives us the following identity: 
\beq\label{6-11}
{\tilde c}_{\alpha}={\tilde c}'_{\alpha}\ , \qquad
s_{\alpha}{\tilde c}_{\ovl \alpha}=s_{\alpha}{\tilde c}'_{\ovl \alpha}\ . 
\eeq
The above indicates that the original fermion operators ${\tilde c}_{\alpha}$, 
${\tilde c}_{\alpha}^*$, $s_{\alpha}{\tilde c}_{\ovl \alpha}$ and $s_{\alpha}{\tilde c}_{\ovl \alpha}^*$ 
are connected with ${\tilde b}_{\alpha}$, ${\tilde b}_{\alpha}^*$, $s_{\alpha}{\tilde b}_{\ovl \alpha}$, 
$s_{\alpha}{\tilde b}_{\ovl \alpha}^*$, ${\hat A}$ and ${\hat A}^*$ through the relations 
(\ref{6-9}) and (\ref{6-11}). 
If ${\hat A}/\sqrt{2{\wtilde S}}$ and ${\hat A}^*/\sqrt{2{\wtilde S}}$ can be 
regarded as $c$-numbers, the relation (\ref{6-9}) ((\ref{6-11})) are reduced to the 
Bogoliubov transformation and ${\tilde b}_{\alpha}$ etc. becomes the 
quasi-particle operators.

With the use of the relations (\ref{6-9}) and (\ref{6-11}), we have the following relation: 
\beq\label{6-12}
& &{\hat S}_+={\hat A}^*\cdot\sqrt{2{\wtilde S}-{\hat A}^*{\hat A}} \ , \qquad
{\hat S}_-=\sqrt{2{\wtilde S}-{\hat A}^*{\hat A}}\cdot {\hat A} \ , \nonumber\\
& &{\hat S}_0={\hat A}^*{\hat A}-{\wtilde S}\ . 
\eeq
Here, ${\wtilde S}$ is defined in the relation (\ref{6-4}). 
With the use of the relation (\ref{6-12}), 
${\hat {\mib S}}^2$ can be calculated as 
\beq\label{6-13}
{\hat {\mib S}}^2={\wtilde S}({\wtilde S}+1) \ . 
\eeq
Therefore, ${\wtilde S}$ indicates the operator for the magnitude of the $S$-spin. 
On the other hand, ${\wtilde R}_{\pm,0}$ can be expressed as 
\beq\label{6-14}
& &{\wtilde R}_+=\sum_{\alpha}{\tilde b}_{\alpha}^*{\tilde b}_{\ovl \alpha}\ , \qquad
{\wtilde R}_-=\sum_{\alpha}{\tilde b}_{\ovl \alpha}^*{\tilde b}_{\alpha}\ , \nonumber\\
& &{\wtilde R}_0=\frac{1}{2}\sum_{\alpha}({\tilde b}_{\alpha}^*{\tilde b}_{\alpha}
-{\tilde b}_{\ovl \alpha}^*{\tilde b}_{\ovl \alpha})\ . 
\eeq
It may be interesting, but natural that the seniority algebra can be expressed in terms of the quasi-fermions. 
It does not depend on $({\hat A},\ {\hat A}^*)$. 
Further, for the magnitude of the $R$-spin, we have 
\beq\label{6-15}
{\wtilde R}=\Omega_0-{\wtilde S} \ , \qquad
{\wtilde R}=\frac{1}{2}\sum_{\alpha}({\tilde b}_{\alpha}^*{\tilde b}_{\alpha}
+{\tilde b}_{\ovl \alpha}^*{\tilde b}_{\ovl \alpha})\ . 
\eeq
On the other hand, ${\hat S}_{\pm,0}$ are expressed in terms of ${\hat A}$ and ${\hat A}^*$ and 
through ${\wtilde S}=\Omega_0+{\wtilde R}$, they depend 
on ${\tilde b}_{\alpha}^*$ etc.
It is in the same situation as that in the case of the boson realization. 
The comparison of the forms (\ref{6-12}) and (\ref{6-13}) with the relations (\ref{2-14a}) 
and (\ref{2-20}) leads us to the following: 
\bsub\label{6-16}
\beq
& &{\hat d}_P^*{\hat d}_{\ovl P} \sim \sum_{\alpha}{\tilde b}_{\alpha}^*{\tilde b}_{\ovl \alpha} \ , \qquad
{\hat d}_{\ovl P}^*{\hat d}_{P} \sim \sum_{\alpha}{\tilde b}_{\ovl \alpha}^*{\tilde b}_{\alpha} \ ,
\nonumber\\
& &\frac{1}{2}({\hat d}_P^*{\hat d}_P-{\hat d}_{\ovl P}^*{\hat d}_{\ovl P})
\sim
\frac{1}{2}\sum_{\alpha}({\tilde b}_{\alpha}^*{\tilde b}_{\alpha}-{\tilde b}_{\ovl \alpha}^*{\tilde b}_{\ovl \alpha}) \ .
\label{6-16a}
\eeq
Further, we have 
\beq\label{6-16b}
\frac{1}{2}({\hat d}_P^*{\hat d}_P+{\hat d}_{\ovl P}^*{\hat d}_{\ovl P})
\sim
\frac{1}{2}\sum_{\alpha}({\tilde b}_{\alpha}^*{\tilde b}_{\alpha}+{\tilde b}_{\ovl \alpha}^*{\tilde b}_{\ovl \alpha}) \ .
\eeq
\esub
The relation (\ref{6-16}) suggests us that the use of the quasi-fermions permits us to complete our aim mentioned in 
the introductory part of this section. 
At the ending of this section, we add a small remark: 
In order to treat fermions in the frame of classical mechanics, 
Casalbuoni introduced the Grassmann variables,\cite{14} which are quantized in Ref.\citen{12}.

\section{The Lipkin model-Part II}

In this section, we consider some features different from those in \S 5. 
Following the promise in \S 5, we will discuss the Lipkin model in relation to 
the isovector pairing model, which is formulated in terms of the $so(5)$-algebra.\cite{7} 
This model consists of ten generators, in which the fermion-pair type generators are as follows: 
\bsub\label{7-1}
\beq
& &{\wtilde Q}_+^*=\sum_m (-)^{j-m}{\tilde c}_{pm}^*{\tilde c}_{p{\ovl m}}^* \ , \qquad
{\wtilde Q}_+=\sum_m (-)^{j-m}{\tilde c}_{p{\ovl m}}{\tilde c}_{p{m}} \ ,
\label{7-1a}\\
& &{\wtilde Q}_0^*=\sum_m (-)^{j-m}{\tilde c}_{pm}^*{\tilde c}_{n{\ovl m}}^* \ , \qquad
{\wtilde Q}_0=\sum_m (-)^{j-m}{\tilde c}_{n{\ovl m}}{\tilde c}_{p{m}} \ ,
\label{7-1b}\\
& &{\wtilde Q}_-^*=\sum_m (-)^{j-m}{\tilde c}_{nm}^*{\tilde c}_{n{\ovl m}}^* \ , \qquad
{\wtilde Q}_-=\sum_m (-)^{j-m}{\tilde c}_{n{\ovl m}}{\tilde c}_{n{m}} \ .
\label{7-1c}
\eeq
\esub
Other four generators have already appeared in the discussion on the isoscalar pairing model 
as the relations (\ref{4-1}) and (\ref{4-3}): 
\beq
& &{\tilde \tau}_+=\sum_m {\tilde c}_{pm}^*{\tilde c}_{nm} \ , \quad
{\tilde \tau}_-=\sum_m {\tilde c}_{nm}^*{\tilde c}_{pm} \ , \quad
{\tilde \tau}_0=\frac{1}{2}\sum_m ({\tilde c}_{pm}^*{\tilde c}_{pm}
-{\tilde c}_{nm}^*{\tilde c}_{nm}) \ , 
\label{7-2}\\
& &{\wtilde \Sigma}=\frac{1}{2}\sum_m ({\tilde c}_{pm}^*{\tilde c}_{pm}
+{\tilde c}_{nm}^*{\tilde c}_{nm})-\Omega \ . 
\label{7-3}
\eeq
The sets $({\wtilde Q}_{\pm,0}^*)$ and $({\wtilde Q}_{\pm,0})$ form the isovectors with respect to the isospin 
$({\tilde \tau}_{\pm,0})$. 
The Casimir operator of the $so(5)$-algebra, ${\hat \Gamma}_{so(5)}$, is expressed as 
\beq\label{7-4}
{\hat \Gamma}_{so(5)}
&=&
\frac{1}{2}({\wtilde Q}_+^*{\wtilde Q}_+ + {\wtilde Q}_-^*{\wtilde Q}_-)
+{\wtilde Q}_0^*{\wtilde Q}_0+{\tilde \tau}_+{\tilde \tau}_-
+{\wtilde \Sigma}({\wtilde \Sigma}-3)+{\tilde \tau}_0({\tilde \tau}_0-1) \nonumber\\
&=&\frac{1}{2}({\wtilde Q}_+^*{\wtilde Q}_+ + {\wtilde Q}_-^*{\wtilde Q}_-)
+{\wtilde Q}_0^*{\wtilde Q}_0+{\tilde \tau}_-{\tilde \tau}_+
+{\wtilde \Sigma}({\wtilde \Sigma}-3)+{\tilde \tau}_0({\tilde \tau}_0+1) \ .
\eeq
The $su(2)$-generators which commute with the above ten generators are as follows: 
\beq\label{7-5}
{\wtilde R}_{\pm,0}={\wtilde R}_{\pm,0}(p)+{\wtilde R}_{\pm,0}(n)\ , 
\eeq
\vspace{-0.6cm}
\bsub\label{7-6}
\beq
& &{\wtilde R}_+(p)=\sum_{m>0}{\tilde c}_{pm}^*{\tilde c}_{p{\ovl m}} \ , \qquad
{\wtilde R}_-(p)=\sum_{m>0}{\tilde c}_{p{\ovl m}}^*{\tilde c}_{p{m}} \ , \nonumber\\
& &
{\wtilde R}_0(p)=\frac{1}{2}\sum_{m>0}({\tilde c}_{pm}^*{\tilde c}_{pm}-
{\tilde c}_{p{\ovl m}}^*{\tilde c}_{p{\ovl m}}) \ , 
\label{7-6a}\\
& &{\wtilde R}_+(n)=\sum_{m>0}{\tilde c}_{nm}^*{\tilde c}_{n{\ovl m}} \ , \qquad
{\wtilde R}_-(n)=\sum_{m>0}{\tilde c}_{n{\ovl m}}^*{\tilde c}_{n{m}} \ , \nonumber\\
& &{\wtilde R}_0(n)=\frac{1}{2}\sum_{m>0}({\tilde c}_{nm}^*{\tilde c}_{nm}-
{\tilde c}_{n{\ovl m}}^*{\tilde c}_{n{\ovl m}}) \ . 
\label{7-6b}
\eeq
\esub
The above $su(2)$-generators are copied from the relation (\ref{4-7}).

We can see that the set (${\wtilde Q}_0^*,\ {\wtilde Q}_0,\ {\wtilde \Sigma})$ forms the 
$su(2)$-algebra and under the following reading, this set is reduced to the Lipkin model 
shown in the relations (\ref{5-4}) and (\ref{5-5}): 
\beq\label{7-7}
n \rightarrow h\ , \quad
{\wtilde Q}_0^* \rightarrow {\wtilde S}_+ \ , \quad
{\wtilde Q}_0 \rightarrow {\wtilde S}_- \ , \quad
{\wtilde \Sigma} \rightarrow {\wtilde S}_0 \ , \quad
{\tilde \tau}_0 \rightarrow \frac{{\wtilde N}}{2}-\Omega \ .
\eeq
Therefore, it is possible to formulate the Lipkin model as the sub-algebra of the 
$so(5)$-algebra which describes the isovector pairing model.

Let us search the minimum weight state for the $so(5)$-algebra. 
For this aim, we set up the relations 
\beq
& &{\wtilde Q}_{\pm,0}\rket{m_0}=0\ , \qquad
{\wtilde R}_-\rket{m_0}=0 \ , 
\label{7-8}\\
& &{\wtilde \Sigma}\rket{m_0}=-s\rket{m_0}\ , \qquad
{\wtilde R}_0\rket{m_0}=-r\rket{m_0} \ . 
\label{7-9}
\eeq
Concerning the isospin, we treat two cases separately: 
\bsub\label{7-10}
\beq
& &{\rm case}\ (1)\ ; \ 
{\tilde \tau}_-\rket{m_0}=0\ , \qquad
{\tilde \tau}_0\rket{m_0}=-\tau\rket{m_0}\ , 
\label{7-10a}\\
& &{\rm case}\ (2)\ ; \ 
{\tilde \tau}_+\rket{m_0}=0\ , \qquad
{\tilde \tau}_0\rket{m_0}=+\tau\rket{m_0}\ , 
\label{7-10b}
\eeq
\esub
With the use of the expressions (\ref{7-1})$\sim$(\ref{7-3}) and (\ref{7-5}) with (\ref{7-6}), 
we obtain the following form for $\rket{m_0}$:
\beq
\label{7-11}
\rket{m_0}=
\left\{
\begin{array}{c}
\displaystyle 
\rket{r,\tau;(pn{\ovl \mu},n{\ovl m})}
=\left(\prod_{j=1,\mu_j>0}^{r-\tau}{\tilde c}_{p{\ovl \mu}_j}^*{\tilde c}_{n{\ovl \mu}_j}^*\right)\!\!
\left(\prod_{i=1,m_i>0}^{2\tau}{\tilde c}_{n{\ovl m}_i}^*\right)\rket{0} \ , \\
\displaystyle 
\rket{r,\tau;(np{\ovl \mu},p{\ovl m})}
=\left(\prod_{j=1,\mu_j>0}^{r-\tau}{\tilde c}_{n{\ovl \mu}_j}^*{\tilde c}_{p{\ovl \mu}_j}^*\right)\!\!
\left(\prod_{i=1,m_i>0}^{2\tau}{\tilde c}_{p{\ovl m}_i}^*\right)\rket{0} \ . \\
\end{array}
\right.
\eeq
Here, the upper and the lower form in the relation (\ref{7-11}) are obtained for the cases (1) and (2), 
respectively. 
The symbol $(pn{\ovl \mu},n{\ovl m})$ denotes the configuration 
$p{\ovl \mu}_1$, $p{\ovl \mu}_2,\cdots$, $p{\ovl \mu}_{r-\tau}$, 
$n{\ovl \mu}_1$, $n{\ovl \mu}_2,\cdots$, $n{\ovl \mu}_{r-\tau}$, 
$n{\ovl m}_1$, $n{\ovl m}_2,\cdots$, $n{\ovl m}_{2\tau}$ and 
$(np{\ovl \mu},p{\ovl m})$ is given by exchanging $p$ and $n$ in $(pn{\ovl \mu},n{\ovl m})$. 
Further, we have 
\beq
& &r+s=\Omega\ , \qquad \tau\leq r \ , 
\label{7-12}\\
& &{\rm the\ eigenvalue\ of}\ \ {\wtilde \Gamma}_{so(5)}=s(s+3)+\tau(\tau+1)\ . 
\label{7-13}
\eeq
We can learn that $2r$ and $\tau$ indicate the seniority number and the 
reduced isospin which characterize the $so(5)$-algebra. 
Therefore, by operating $({\wtilde R}_+)^{r+r_0}$ on the state (\ref{7-11}), 
the minimum weight state of the $so(5)$-algebra is obtained: 
\bsub\label{7-14}
\beq
& &\rket{rr_0,\tau;(pn{\ovl \mu},n{\ovl m})}
=({\wtilde R}_+)^{r+r_0}\rket{r,\tau;(pn{\ovl \mu},n{\ovl m})}\quad
{\rm for\ the\ case\ (1)}\ , 
\label{7-14a}\\
& &\rket{rr_0,\tau;(np{\ovl \mu},p{\ovl m})}
=({\wtilde R}_+)^{r+r_0}\rket{r,\tau;(np{\ovl \mu},p{\ovl m})}\quad
{\rm for\ the\ case\ (2)}\ . 
\label{7-14b}
\eeq
\esub
The orthogonal set for the $so(5)$-algebra is constructed by operating 
${\wtilde Q}_{\pm,0}^*$ and ${\tilde \tau}_+$ (for the state (\ref{7-14a})) 
and ${\tilde \tau}_-$ (for the state (\ref{7-14b})) on the 
state (\ref{7-14}). 
Totally, it is specified by six quantum numbers except the extra quantum numbers 
specifying the minimum weight state, for example, such as $r_0$. 
In this paper, we omit the concrete procedure for this construction, because 
we are interested in the sub-algebra, i.e., the $su(2)$-algebra.

On the basis of the above results on the $so(5)$-algebra, we will discuss the Lipkin model 
under the reading (\ref{7-7}). 
It may be self-evident that the states (\ref{7-14a}) and (\ref{7-14b}) are the 
minimum weight states for the algebra $({\wtilde S}_{\pm,0})$. 
The eigenvalues of ${\wtilde S}_0$ and ${\wtilde R}_0$ are given by $-s$ and $-r$, 
respectively, which are related to each other under the relation (\ref{7-12}) and, then, 
the physical meanings are the same as those given in \S 5. 
The states (\ref{7-14a}) and (\ref{7-14b}) are 
also the eigenstates of ${\tilde \tau}_0$ with the eigenvalues 
$-\tau$ and $+\tau$, respectively. 
As is shown in the relation (\ref{7-7}), total fermion number operator 
${\wtilde N}$ is expressed as ${\wtilde N}=2\Omega+2{\tilde \tau}_0$. 
Therefore, for the states (\ref{7-14a}) and (\ref{7-14b}), the fermion numbers $N$ are 
given by $N=2\Omega-2\tau$ and $N=2\Omega+2\tau$, respectively. 
This implies that ${\tilde \tau}$ plays the same role as that of ${\wtilde R}_0$ in \S 5. 
Then, the role of $({\wtilde R}_{\pm,0})$ in the present case may be interesting.

The states (\ref{7-14a}) and (\ref{7-14b}) $(r_0=-r)$ are expressed in terms of the operators 
$({\tilde c}_{p{\ovl m}}^*,\ {\tilde c}_{n{\ovl m}}^*)$, where ${\ovl m}<0$, 
i.e., $m>0$. 
By operating ${\wtilde R}_+$ successively, these states change their structures and they 
are expressed not only by $({\tilde c}_{p{\ovl m}}^*,\ {\tilde c}_{n{\ovl m}}^*)$ 
but also $({\tilde c}_{p{m}}^*,\ {\tilde c}_{n{m}}^*)$. 
Finally, at $r_0=r$, the minimum weight states are expressed only in terms of 
$({\tilde c}_{p{m}}^*,\ {\tilde c}_{n{m}}^*)$. 
Therefore, for example, at $r_0=0$ which appears in the case 
$r=$even, the state (\ref{7-14}) contains 
$({\tilde c}_{p{\ovl m}}^*,\ {\tilde c}_{n{\ovl m}}^*)$ and 
$({\tilde c}_{p{m}}^*,\ {\tilde c}_{n{m}}^*)$ in equal weight. 
In the case $r=$odd, the situation similar to the case $r=$even appears 
at $r_0=\pm 1/2$. 
We observe these features in the pairing model given in \S 3. 
The minimum weight state in \S 5 does not contain such a distinction. 
In this sense, the minimum weight states in \S 5 and \S 7 may be equivalent to each other, 
but, the state in \S 7 contains the information other than the state in \S 5.

The idea discussed in the above suggests us to formulate the pairing model 
in \S 3 in the present framework. 
We note that the operators (${\wtilde Q}_-^*,\ {\wtilde Q}_-,\ 
({\wtilde \Sigma}-{\tilde \tau}_0-\Omega)/2)$ form the $su(2)$-algebra. 
If these operators read $({\wtilde S}_+,\ {\wtilde S}_-,\ {\wtilde S}_0)$, respectively, 
and the $p$-level is vacant, it is reduced to the pairing model in \S 3. 
If $\tau$ is equal to $r$ in the state given in the upper of (\ref{7-11}), the 
$p$-level becomes vacant and this case corresponds to the minimum weight state of the pairing 
model.

Now, with the aid of the reading (\ref{7-7}), we consider the boson realization of the Lipkin model 
based on the isovector pairing model. 
First, we introduce the counterparts of ${\wtilde R}_{\pm,0}$ and ${\wtilde S}_{\pm,0}$ 
in the boson space. 
The counterpart of ${\wtilde R}_{\pm,0}$ shown in the relation (\ref{7-5}) is 
given in the Schwinger boson representation:
\beq\label{7-15}
{\hat R}_{\pm,0}={\hat R}_{\pm,0}(p)+{\hat R}_{\pm,0}(h)\ , 
\qquad\qquad\qquad\qquad\qquad\qquad
\eeq
\vspace{-0.6cm}
\bsub\label{7-16}
\beq
& &{\hat R}_+(p)={\hat d}_P^*(p){\hat d}_{\ovl P}(p) \ , \qquad
{\hat R}_-(p)={\hat d}_{\ovl P}^*(p){\hat d}_{P}(p) \ , 
\nonumber\\
& &{\hat R}_0(p)=\frac{1}{2}({\hat d}_P^*(p){\hat d}_P(p)
-{\hat d}_{\ovl P}^*(p){\hat d}_{\ovl P}(p))\ , 
\label{7-16a}\\
& &{\hat R}_+(h)={\hat d}_P^*(h){\hat d}_{\ovl P}(h) \ , \qquad
{\hat R}_-(h)={\hat d}_{\ovl P}^*(h){\hat d}_{P}(h) \ , 
\nonumber\\
& &{\hat R}_0(h)=\frac{1}{2}({\hat d}_P^*(h){\hat d}_P(h)
-{\hat d}_{\ovl P}^*(h){\hat d}_{\ovl P}(h))\ , 
\label{7-16b}
\eeq
The magnitudes of the $R_p$- and the $R_h$-spin are given by 
\beq\label{7-16c}
& &{\hat R}(p)=\frac{1}{2}({\hat d}_P^*(p){\hat d}_P(p)
+{\hat d}_{\ovl P}^*(p){\hat d}_{\ovl P}(p))\ , \quad
{\hat R}(h)=\frac{1}{2}({\hat d}_P^*(h){\hat d}_P(h)
+{\hat d}_{\ovl P}^*(h){\hat d}_{\ovl P}(h))\ . \nonumber\\
& &
\eeq
\esub
The case of the $S$-spin is expressed as 
\bsub\label{7-17}
\beq\label{7-17a}
{\hat S}_+={\hat a}^*{\hat b}\ , \qquad
{\hat S}_-={\hat b}^*{\hat a} \ , \qquad
{\hat S}_0=\frac{1}{2}({\hat a}^*{\hat a}-{\hat b}^*{\hat b}) \ . 
\eeq
The magnitude of the $S$-spin is given by 
\beq\label{7-17b}
{\hat S}=\frac{1}{2}({\hat a}^*{\hat a}+{\hat b}^*{\hat b}) \ .
\eeq
\esub
Here, $({\hat d}_P(p),\ {\hat d}_{P}^*(p))$ etc. denote boson operators. 
In order to get the counterpart of the state $\rket{m_0}$ 
shown in the relation (\ref{7-11}), we must investigate the coupling scheme 
of the $R_{p}$- and the $R_{h}$-spin governing the state $\rket{m_0}$. 
The connection of the $S$-spin with the $R$-spin is given by the 
relation (\ref{7-12}).

The state $\rket{m_0}$ given in the relation (\ref{7-11}) satisfies the relation 
\bsub\label{7-18}
\beq
& &{\wtilde R}_-(p)\rket{m_0}={\wtilde R}_-(h)\rket{m_0}=0\ , \quad
{\rm i.e.,}\quad
{\wtilde R}_-\rket{m_0}=0\ , 
\label{7-18a}\\
& &{\wtilde R}_0(p)\rket{m_0}=-r_p\rket{m_0}\ , \qquad
{\wtilde R}_0(h)\rket{m_0}=-r_h\rket{m_0}\ , \nonumber\\
& &\ \ 
{\rm i.e.,}\quad
{\wtilde R}_0\rket{m_0}=-(r_p+r_h)\rket{m_0}\ . 
\label{7-18b}
\eeq
\esub
Here, $r_p$ and $r_h$ are given as 
\bsub\label{7-19}
\beq
& &r_p=\frac{1}{2}(r-\tau)\ , \qquad
r_h=\frac{1}{2}(r+\tau)\quad
{\rm for\ the\ upper\ state\ of}\ (7\!\cdot\! 11) \ , 
\label{7-19a}\\
& &r_p=\frac{1}{2}(r+\tau)\ , \qquad
r_h=\frac{1}{2}(r-\tau)\quad
{\rm for\ the\ lower\ state\ of}\ (7\!\cdot\! 11) \ . 
\label{7-19b}
\eeq
\esub
The relation (\ref{7-19}) leads us to 
\beq\label{7-20}
r_p+r_h=r\ , \quad {\rm i.e.,}\quad
(r_p+r_h)(r_p+r_h+1)=r(r+1) \ . 
\eeq
The above discussion gives us the following picture for the coupling scheme: 
The directions of both spins are the same as each other.

Under the above preparation, we will investigate the other kind of the boson 
realization for the Lipkin model. 
First, we notice the following operator identity: 
\beq\label{7-21}
{\hat {\mib R}}^2={\hat {\mib T}}^2\ , \qquad
{\hat {\mib T}}^2=-{\hat T}_+{\hat T}_-+{\hat T}_0({\hat T}_0-1) \ . 
\eeq
Here, ${\hat {\mib R}}^2$ denotes the Casimir operator of the $su(2)$-algebra, the 
generators of which are defined in the relations (\ref{7-15}) and (\ref{7-16}). 
The operator ${\hat {\mib T}}^2$ denotes the Casimir operator of the $su(1,1)$-algebra, 
the generators of which are defined as follows: 
\bsub\label{7-22}
\beq
& &{\hat T}_+={\hat d}_P^*(p){\hat d}_{\ovl P}^*(h)
-{\hat d}_P^*(h){\hat d}_{\ovl P}^*(p) \ , \qquad
{\hat T}_-={\hat d}_{\ovl P}(h){\hat d}_{P}(p)
-{\hat d}_{\ovl P}(p){\hat d}_{P}(h) \ , 
\nonumber\\
& &{\hat T}_0=\frac{1}{2}
({\hat d}_P^*(p){\hat d}_P(p)+{\hat d}_{\ovl P}^*(p){\hat d}_{\ovl P}(p)
+{\hat d}_P^*(h){\hat d}_P(h)+{\hat d}_{\ovl P}^*(h){\hat d}_{\ovl P}(h))+1\ , 
\label{7-22a}
\eeq
The set $({\hat T}_{\pm,0})$ satisfies 
\beq\label{7-22b}
[\ {\hat T}_+\ , \ {\hat T}_-\ ]=-2{\hat T}_0 \ , \qquad
[\ {\hat T}_0 \ , \ {\hat T}_{\pm}\ ]=\pm {\hat T}_{\pm}\ . 
\eeq
\esub
Detailed argument can be found in the paper by the present authors 
(J. da P. \& M. Y.) together with Kuriyama.\cite{15} 
It may be interesting to see that ${\hat T}_0$ can be expressed in terms of the 
operators for the magnitudes of the $R_{p}$- and the $R_{h}$-spins: 
\beq\label{7-23}
& &{\hat T}_0={\hat R}(p)+{\hat R}(h) \ , \nonumber\\
& &{\rm i.e.,}\quad {\hat T}_0({\hat T}_0-1)
=({\hat R}(p)+{\hat R}(h))({\hat R}(p)+{\hat R}(h)+1) \ . 
\eeq
The term $({\hat R}(p)+{\hat R}(h))$ indicates the simple sum of the magnitudes 
of the two $su(2)$-spins. 
Let the eigenstate of ${\hat R}(p)$, ${\hat R}(h)$, ${\hat {\mib R}}^2$ and 
${\hat R}_0$, $\ket{\lambda;r_pr_hrr_0}$, satisfy the relation 
\beq\label{7-24}
{\hat T}_-\ket{\lambda;r_pr_hrr_0}=0\  \qquad
\bra{\lambda;r_pr_hrr_0}{\hat T}_+=0\ . 
\eeq
Then, for the state $\ket{\lambda;r_pr_hrr_0}$, we have 
\beq\label{7-25}
r(r+1)=(r_p+r_h)(r_p+r_h+1) \ . 
\eeq
The relation (\ref{7-24}) is nothing but the result (\ref{7-20}) and, then, 
the condition (\ref{7-23}) presents us the picture that the directions of the $R_{p}$- 
and the $R_{h}$-spin are the same as each other.

The relation (\ref{7-24}) suggests us to adopt the idea presented by Dirac 
for the constraint system. 
In \S 6, we have adopted this idea for the case of many-fermion system. 
We require the following constraints: 
\beq\label{7-26}
{\hat T}_- \approx 0 \ , \qquad
{\hat T}_+\approx 0 \ . 
\eeq
Then, we define the Dirac bracket for ${\hat A}$ and ${\hat B}$, which is denoted as 
$[[\ {\hat A}\ , \ {\hat B}\ ]]$, in the form
\bsub\label{7-27}
\beq\label{7-27a}
[[\ {\hat A}\ , \ {\hat B}\ ]]=[\ {\hat A} \ , \ {\hat B}\ ]
&+&[\ {\hat A} \ , \ {\hat T}_+\ ]\left(
[\ {\hat T}_+\ , \ {\hat T}_-\ ]\right)^{-1} [\ {\hat T}_-\ , \ {\hat B}\ ] 
\nonumber\\
&+&[\ {\hat A}\ , \ {\hat T}_- \ ]\left([\ {\hat T}_- \ , \ {\hat T}_+\ ]\right)^{-1}
[\ {\hat T}_+\ , \ {\hat B}\ ] \ .
\eeq
By using the ordinary boson commutation relations, 
for example, such as\break 
$[\ {\hat d}_P(p)\ , \ {\hat d}_P^*(p)\ ]=1$, 
we calculate the right-hand side of the relation (\ref{7-27a}). 
Consequently, we can express $[[\ {\hat A}\ , \ {\hat B}\ ]]$ as a function of bosons 
${\hat d}_P^*(p)$ etc., which is denoted as ${\hat C}_{AB}$, namely, we have 
\beq\label{7-27b}
[[\ {\hat A}\ , \ {\hat B}\ ]]={\hat C}_{AB} \ .
\eeq
\esub 
Under the above result, we set up the commutation relation for ${\hat A}$ and ${\hat B}$ 
in the following form: 
\beq\label{7-28}
[\ {\hat A}\ , \ {\hat B}\ ]={\hat C}_{AB}\ .
\eeq
For example, we have 
\beq\label{7-29}
[\ {\hat d}_P(p)\ , \ {\hat d}_P^*(p)\ ]=1-
{\hat d}_{\ovl P}^*(h)\left[
2({\hat R}(p)+{\hat R}(h)+1)\right]^{-1}
{\hat d}_{\ovl P}(h)\ . 
\eeq
For the $R$-spin, we adopt the Schwinger boson representation shown in the forms 
(\ref{7-15}) and (\ref{7-16}) with the commutation relations (\ref{7-29}), etc. 
For the $S$-spin, we adopt the Holstein-Primakoff representation in the following form: 
\beq\label{7-30}
& &{\hat S}_+={\hat A}^*\cdot\sqrt{2(\Omega-({\hat R}(p)+{\hat R}(h)))
-{\hat A}^*{\hat A}}\ , \nonumber\\
& &{\hat S}_-=\sqrt{2(\Omega-({\hat R}(p)+{\hat R}(h)))
-{\hat A}^*{\hat A}}\cdot{\hat A}\ , \nonumber\\
& &{\hat S}_0={\hat A}^*{\hat A}-(\Omega-({\hat R}(p)+{\hat R}(h)))\ . 
\eeq
Of course, we used the relation 
\beq\label{7-31}
{\hat S}=\Omega-{\hat R}\ , \qquad
{\hat R}={\hat R}(p)+{\hat R}(h) \ . 
\eeq
Thus, we obtained the boson realization of the Lipkin model which is 
closely related to the isovector pairing model.

\section{Concluding remarks}

In this paper, we formulated the $su(2)$-algebraic many-fermion model 
in rather general scheme. 
In our idea, the $su(2)$-algebra $({\wtilde R}_{\pm,0})$, which we called the 
auxiliary algebra, plays a decisive role for determining the minimum weight state. 
Through the use of this algebra, we can learn various aspects of the models, 
some of which are newly derived. 
Further, we showed that the idea adopted in the $su(2)$-algebra is also applicable to the 
$so(5)$-algebra which treats the isovector pairing model and the Lipkin model 
is formulated as a sub-algebra under the reading (\ref{7-7}).

Finally, we will give a small comment. 
In the relations (\ref{2-1}) and (\ref{2-5}), we make the following replacement: 
\beq\label{8-1}
{\tilde c}_\alpha^*={\tilde \gamma}_{+\alpha}^* \ , \qquad
s_\alpha {\tilde c}_{\ovl \alpha}^*={\tilde \gamma}_{-\alpha} \ .
\eeq
The operators $({\tilde \gamma}_{+\alpha} \ , \ {\tilde \gamma}_{+\alpha}^*)$ and 
$({\tilde \gamma}_{-\alpha} \ , \ {\tilde \gamma}_{-\alpha}^*)$ are also fermions. 
The sets $({\wtilde S}_{\pm,0})$ and $({\wtilde R}_{\pm,0})$ can be rewritten in the form 
\beq
& &{\wtilde S}_+=\sum_{\alpha}{\tilde \gamma}_{+\alpha}^*{\tilde \gamma}_{-\alpha}\ , 
\qquad
{\wtilde S}_-=\sum_{\alpha}{\tilde \gamma}_{-\alpha}^*{\tilde \gamma}_{+\alpha}\ , 
\qquad
{\wtilde S}_0=\frac{1}{2}\sum_{\alpha}({\tilde \gamma}_{+\alpha}^*{\tilde \gamma}_{+\alpha}
-{\tilde \gamma}_{-\alpha}^*{\tilde \gamma}_{-\alpha}) , 
\label{8-2}\\
& &{\wtilde R}_+=\sum_{\alpha}s_{\alpha}{\tilde \gamma}_{+\alpha}^*{\tilde \gamma}_{-\alpha}^*\ , 
\quad
{\wtilde R}_-=\sum_{\alpha}s_{\alpha}{\tilde \gamma}_{-\alpha}{\tilde \gamma}_{+\alpha}\ , 
\quad
{\wtilde R}_0=\frac{1}{2}\sum_{\alpha}({\tilde \gamma}_{+\alpha}^*{\tilde \gamma}_{+\alpha}
+{\tilde \gamma}_{-\alpha}^*{\tilde \gamma}_{-\alpha})-\Omega_0 . \nonumber\\
& &\label{8-3}
\eeq
For the forms (\ref{8-2}) and (\ref{8-3}), let us adopt the reading 
\beq\label{8-4}
{\tilde \gamma} \rightarrow {\tilde c} \ , \qquad
{\tilde \gamma}^* \rightarrow {\tilde c}^*\ , \qquad
+\alpha \rightarrow \alpha \ , \qquad 
-\alpha \rightarrow {\ovl \alpha}\ . 
\eeq
Then, it can be seen that the roles of $({\wtilde S}_{\pm,0})$ and $({\wtilde R}_{\pm,0})$ 
mentioned in \S 2 are reversed. 
Consequently, with the aid of the auxiliary algebra shown in the 
form (\ref{8-3}), we can describe the $su(2)$-algebraic model 
expressed in terms of the form (\ref{8-2}). 
For example, the Lipkin model can be described without introducing the particle and hole operators 
shown in the relation (\ref{5-3}). 
The above is our small comment.


\section*{Acknowledgement}

One of the authors (Y.T.) 
is partially supported by the Grants-in-Aid of the Scientific Research 
(No.23540311) from the Ministry of Education, Culture, Sports, Science and 
Technology in Japan.



\end{document}